\documentclass[english,superscriptaddress,floatfix,twocolumn]{revtex4}
\usepackage[T1]{fontenc}
\usepackage[latin9]{inputenc}
\usepackage[pdftex]{xcolor}
\usepackage{pdfcolmk}
\usepackage{babel}
\usepackage{mathrsfs}
\usepackage{bm}
\usepackage{amsmath}
\usepackage{amssymb}
\usepackage[pdftex]{graphicx}
\usepackage{esint}
\PassOptionsToPackage{normalem}{ulem}
\usepackage{ulem}
\usepackage[unicode=true,
 bookmarks=false,
 breaklinks=true,pdfborder={0 0 1},backref=section,colorlinks=true]
 {hyperref}
\hypersetup{
 pdfcreator={},pdfproducer={LaTeX with hyperref},linkcolor=blue,anchorcolor=blue,citecolor=blue,filecolor=red,menucolor=red,pagecolor=red,urlcolor=blue,pdfstartview=FitV,pdfhighlight=/I,pdfpagelayout=OneColumn,hypertexnames=true}

\makeatletter

\providecolor{lyxadded}{rgb}{0,0,1}
\providecolor{lyxdeleted}{rgb}{1,0,0}

\@ifundefined{textcolor}{}
{%
 \definecolor{BLACK}{gray}{0}
 \definecolor{WHITE}{gray}{1}
 \definecolor{RED}{rgb}{1,0,0}
 \definecolor{GREEN}{rgb}{0,1,0}
 \definecolor{BLUE}{rgb}{0,0,1}
 \definecolor{CYAN}{cmyk}{1,0,0,0}
 \definecolor{MAGENTA}{cmyk}{0,1,0,0}
 \definecolor{YELLOW}{cmyk}{0,0,1,0}
}

\usepackage{babel}
\usepackage{upgreek}

 \@ifundefined{textcolor}{}{%
  \definecolor{BLACK}{gray}{0}
  \definecolor{WHITE}{gray}{1}
  \definecolor{RED}{rgb}{1,0,0}
  \definecolor{GREEN}{rgb}{0,1,0}
  \definecolor{BLUE}{rgb}{0,0,1}
  \definecolor{CYAN}{cmyk}{1,0,0,0}
  \definecolor{MAGENTA}{cmyk}{0,1,0,0}
  \definecolor{YELLOW}{cmyk}{0,0,1,0}
 }

 \@ifundefined{textcolor}{}{
  \definecolor{BLACK}{gray}{0}
  \definecolor{WHITE}{gray}{1}
  \definecolor{RED}{rgb}{1,0,0}
  \definecolor{GREEN}{rgb}{0,1,0}
  \definecolor{BLUE}{rgb}{0,0,1}
  \definecolor{CYAN}{cmyk}{1,0,0,0}
  \definecolor{MAGENTA}{cmyk}{0,1,0,0}
  \definecolor{YELLOW}{cmyk}{0,0,1,0}
 }
 \@ifundefined{definecolor}{
 }{}
 \@ifundefined{definecolor}{color}{}

\usepackage{color}

\newcommand{\be}{\begin{equation}}
\newcommand{\ee}{\end{equation}}
\newcommand{\bea}{\begin{eqnarray}}
\newcommand{\eea}{\end{eqnarray}}
\newcommand{\bse}{\begin{subequations}}
\newcommand{\ese}{\end{subequations}}

\usepackage{color}
\definecolor{d_red}{cmyk}{0.00, 0.81, 1.00, 0.27}
\definecolor{d_orange}{cmyk}{0.00, 0.33, 1.00, 0.00}
\definecolor{d_blue}{cmyk}{0.78, 0.47, 0.00, 0.20}
\definecolor{d_lgreen}{cmyk}{0.07, 0.00, 0.79, 0.29}
\definecolor{d_green}{cmyk}{0.66, 0.00, 0.71, 0.56}
\definecolor{d_blue}{cmyk}{0.78, 0.47, 0.00, 0.20}
\definecolor{d_dblue}{cmyk}{0.91, 0.79, 0.00, 0.22}
\definecolor{d_pink}{cmyk}{0.0, 0.79, 0.37, 0.29}
\definecolor{d_purple}{cmyk}{0.16, 0.54, 0.00, 0.70}
\definecolor{d_paleblue}{cmyk}{0.669, 0.338, 0.00, 0.373}
\definecolor{d_dpaleblue}{cmyk}{0.441, 0.290, 0.00, 0.580}
\definecolor{d_brown}{cmyk}{0.0, 0.490, 0.930, 0.350}
\definecolor{d_turquoise}{cmyk}{0.630, 0.04, 0.0, 0.440}
\definecolor{KIT-green}{RGB}{0, 150,130}
\definecolor{KIT-blue}{RGB}{70,100,170}

\def\bmx{\begin{pmatrix}}
\def\emx{\end{pmatrix}}

\usepackage[figure,table]{hypcap}

\makeatother

\begin{document}

\title{The boundary conditions of viscous electron flow}

\author{Egor I. Kiselev}

\affiliation{Institut f\"ur Theorie der Kondensierten Materie, Karlsruher Institut
f\"ur Technologie, 76131 Karlsruhe, Germany}

\author{J\"org Schmalian}

\affiliation{Institut f\"ur Theorie der Kondensierten Materie, Karlsruher Institut
f\"ur Technologie, 76131 Karlsruhe, Germany}

\affiliation{Institut f\"ur Festkörperphysik, Karlsruher Institut für Technologie,
76131 Karlsruhe, Germany}
\begin{abstract}
The sensitivity of charge, heat, or momentum transport to the sample
geometry is a hallmark of viscous electron flow. Therefore hydrodynamic
electronics requires a detailed understanding of electron flow in
finite geometries. The solution of the corresponding generalized Navier-Stokes
equations depends sensitively on the nature of boundary conditions.
The latter can be characterized by a slip length $\zeta$ with extreme
cases being no-slip $\left(\zeta\rightarrow0\right)$ and no-stress
$\left(\zeta\rightarrow\infty\right)$ conditions. We develop a kinetic
theory that determines the temperature dependent slip length at a
rough interface for Dirac liquids, e.g. graphene, and for Fermi liquids.
For strongly disordered edges that scatter electrons in a fully-diffuse
way, we find that the slip length is of the order of the momentum
conserving mean free path $l_{ee}$ that determines the electron viscosity.
For boundaries with nearly specular scattering $\zeta$ is parametrically
large compared to $l_{ee}$. Since for all quantum fluids $l_{ee}$
diverges as $T\rightarrow0$, the ultimate low-temperature flow is
always in the no-stress regime. Only at intermediate $T$ and for
sufficiently large sample sizes can the slip lengths be short enough
such that no-slip conditions are appropriate. We discuss numerical
examples for several experimentally investigated systems. To identify
hydrodynamic flow governed by no-stress boundary conditions, we propse
the transport through an infinitely long strip containing an impenetrable
circular obstacle. 
\end{abstract}
\maketitle

\section{Introduction}

The fluid flow of liquids is governed by the laws of hydrodynamics.
If collisions are sufficiently strong and lead to local thermalization,
yet respect the laws of charge, energy, and momentum conservation,
hydrodynamics should apply \cite{Forster}. Inhomogeneous flow velocity
profiles of the Couette and Poiseuille type, vorticity of flow, or
turbulent flow are among the indicators of hydrodynamic behavior.
Starting with the pioneering work by Gurzhi \cite{Gurzhi1968} in
1968, the theoretical foundations of electron hydrodynamics have been
discussed for a range electronic systems \cite{deJong1995,Kashuba2008,Fritz2008,Mueller2009,Andreev2011,Davison2014,Torre2015,Kashuba2018,Briskot2015,Alekseev2016,Levitov2016,Scaffidi2017,Lucas2018,Narozhny2017,Lucas2016_1,Lucas2016_2,Derek2018}.
For electron hydrodynamics to apply, electron-electron collissions
should dominate. Thus, the temperature should be below $T_{{\rm ph}}$
where electron-phonon scattering starts violating energy and momentum
conservation of the electronic subsystems. At the same time $T$ should
be above $T_{{\rm imp}}$ where impurities dominate, violating momentum
conservation. Only if $T_{{\rm imp}}<T_{{\rm ph}}$ is there a window
for hydrodynamic electronics, explaining the need for ultra-clean
materials.

Examples for recent experimental investigations of hydrodynamic electronics
include the observation of non-local momentum relaxation in the delafossite
PdCoO$_{2}$ \cite{Moll2016} and the Weyl semimetal WP$_{2}$ \cite{Gooth2017},
systems that are special because of their exceptionally low residual
resistivity. In parallel, advances in the fabrication of high-quality
graphene led to the observation of hydrodynamic Coulomb drag \cite{Titov2013},
violations of the Wiedeman Franz law for the thermal transport \cite{Crossno2016}
and the Mott relation for the thermoelectric transport \cite{Ghahari2016},
a negative local resistance due to flow with vorticity \cite{Bandurin},
and superballistic flow \cite{KrishnaKumar2017}. In graphene at the
charge neutrality point and other Dirac systems, electron-hole puddles
form due to disorder effects \cite{Martin2008}. These puddles result
in a local variation of the chemical potential $\delta\mu$. To observe
hydrodynamic effects, $\left|\delta\mu\right|<k_{B}T$ must hold -
a condition achieved in current experiments \cite{Crossno2016}. A
key common feature of all those experiments is the fact that finite
geometries strongly affect the electron flow. In fact the sensitivity
of the flow profile to boundaries has been a key strategy to identify
hydrodynamic flow. 

The theoretical modelling of viscous electron flow is often based
on the solution of kinetic equations \cite{Mueller2009,Titov2013,Briskot2015,Alekseev2016,Levitov2016,Scaffidi2017,Link2018}.
A very efficient description, particularly appropriate for complex
geometries is based on the Navier-Stokes equations for the flow velocity
$\mathbf{u}\left(\mathbf{r},t\right)$. For Lorentz and Galilei-invariant
systems the Navier-Stokes equations are dictated by symmetry \cite{Forster}.
In the general setting, they can be derived from the kinetic equation,
see e.g. Ref. \cite{Mueller2009,Briskot2015}. Not surprisingly, the
solutions of these equations depend sensitively on the imposed conditions
at the sample boundaries. Let $S$ be the boundary of the sample.
Popular boundary conditions are the no-slip condition 
\begin{equation}
\left.u_{\alpha}^{t}\right|_{S}=0,
\end{equation}
where $\mathbf{u}^{t}=\mathbf{u}-\left(\mathbf{u}\cdot\mathbf{n}\right)\mathbf{n}$
is the tangential velocity of a boundary with normal vector\textbf{
$\mathbf{n}$}, and the no-stress condition
\begin{equation}
\left.n_{\beta}\frac{\partial u_{\alpha}^{t}}{\partial x_{\beta}}\right|_{S}=0.\label{eq:no_stress_condition}
\end{equation}
The no-slip condition is the relevant one for most liquid-solid interfaces.
Liquid particles at the surface do not move with the fluid flow, an
effect either explained in terms of surface roughness or due to attractive
interactions between solid and liquid particles, see Ref. \cite{Zhu2002,Lauga2007}.
On the other hand, the liquid flow near a liquid-gas interface is
often characterized by the no-stress condition, i.e. the tangential
stress at the interface is continuous. As discussed by Maxwell \cite{Maxwell1867},
a boundary condition, that includes both cases as limits is 

\begin{equation}
\left.u_{\alpha}^{t}\right|_{S}=\zeta\left.n_{\beta}\frac{\partial u_{\alpha}^{t}}{\partial x_{\beta}}\right|_{S},\label{eq:General_boundary_condition}
\end{equation}
where $\zeta$ is the slip length.
\begin{figure}
\begin{centering}
\includegraphics[angle=270,scale=0.34]{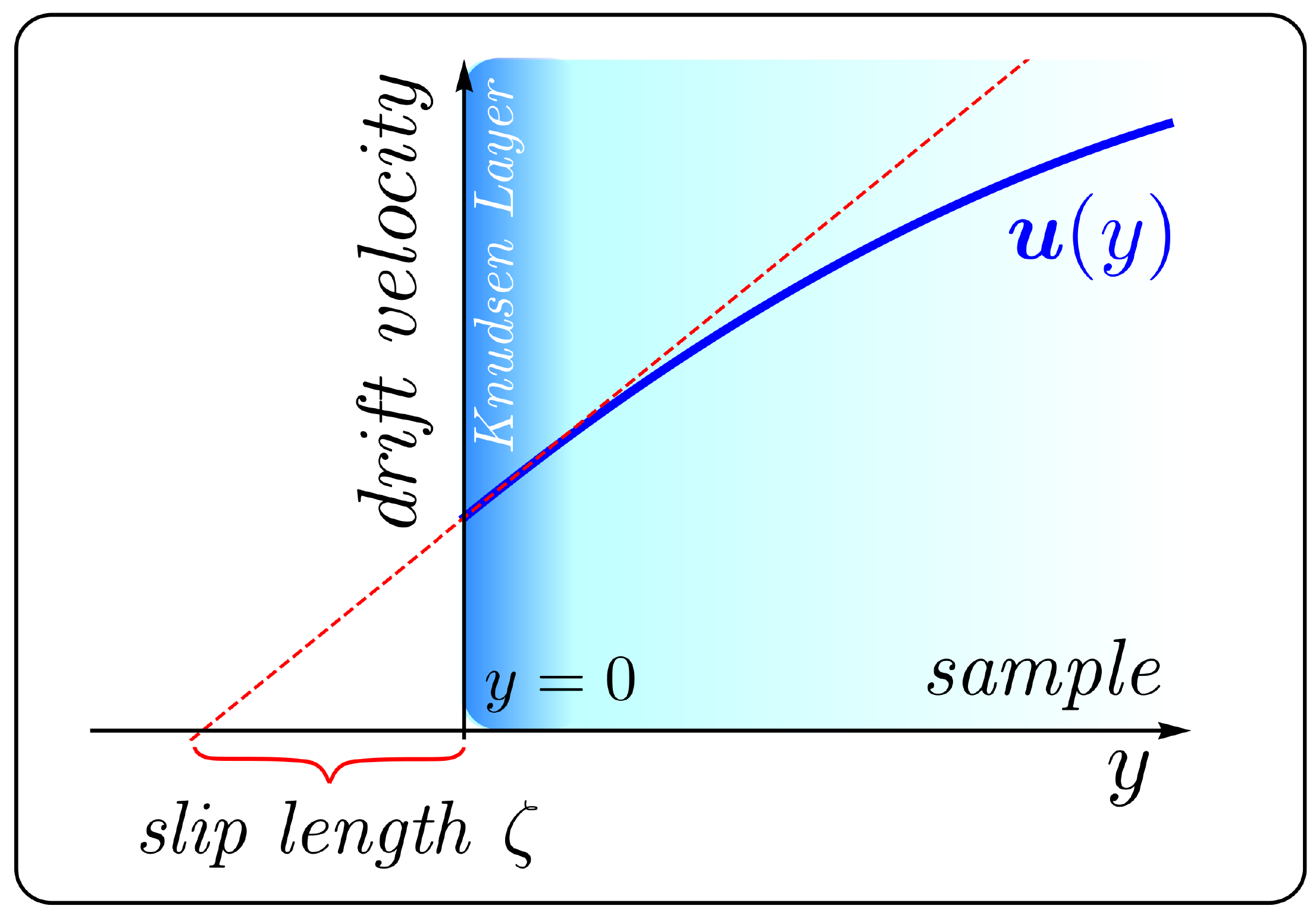}
\par\end{centering}

\caption{Velocity profile $\bm{u}\left(y\right)$ of an electron liquid near
the edge of a sample. The boundary condition for $\bm{u}\left(y\right)$
is given by Eq. (\ref{eq:General_boundary_condition}) so that $u\left(y=0\right)=\zeta\frac{\partial u}{\partial y}$.
The slip length $\zeta$ characterizes the behavior of a liquid near
the edge. It corresponds to the length where the extrapolated velocity
vanishes. Also depicted is the Knudsen layer - an approximately one
mean free path thin layer along the boundary, where the collisions
of particles with the wall are as important as collisions among each
other (see also Fig. \ref{fig:Knudsen-layer}).\label{fig:Slip_length_def}}
\end{figure}
It corresponds to the length where the extrapolated boundary velocity
vanishes (see Fig. \ref{fig:Slip_length_def}). Clearly, $\zeta\rightarrow0$
and $\zeta\rightarrow\infty$ correspond to no-slip and no-stress
conditions, respectively. Since the origin of tangential stress in
a fluid is purely visous, one expects that the slip length is another
quantity that can be determined from kinetic theory, like diffusivities
or viscosities. Indeed, for rarified gases, Maxwell found that $\zeta$
is essentially given by the momentum-conserving mean free path, a
result fully consistent with numerical simulations \cite{Morris1992}.
Other systems where a finite slip length is of relevance are classical
fluids affected by soft hydrodynamic modes \cite{Nieuwoudt1984,Wolynes1976}
polymer melts \cite{deGennes1979}, phononic liquids \cite{Levinson1977}
and $^{3}{\rm He}$ at low temperatures in the normal and superfluid
state \cite{HJensen1980,Einzel1984,Einzel1990,Einzel1997}. Ref. \cite{Delacretaz2017}
reports that in the quantum Hall regime no-stress conditions must
be applied to agree with known results for the quantized Hall conductance. 

\begin{figure}
\includegraphics[angle=270,scale=0.19]{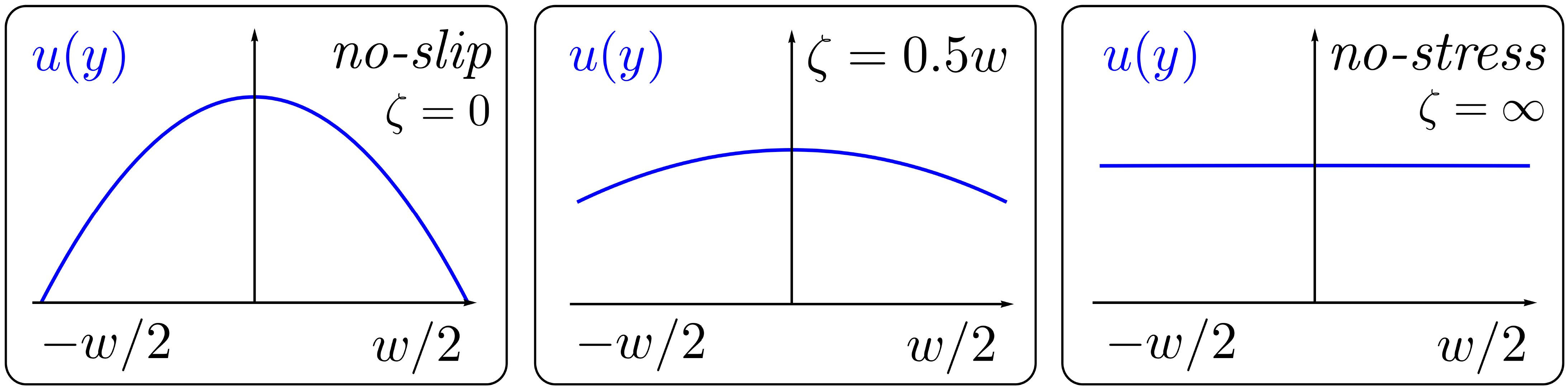}\caption{Flow profiles of a wire of thickness $w$ through which passes a current
$I$. For no-slip boundary conditions a parabolic flow profile, typical
for Poiseuille flow, is realized. With growing slip length the flow
profile becomes flat, i.e. more similar to Ohmic flow. \label{fig:Profiles_Slip_Length}}
\end{figure}

Let us demonstrate the importance of a finite slip length for the
fluid flow for a simple example. Consider the flow of a two-dimensional
system that is governed by the linear, stationary limit of the Navier
Stokes equation. For a strip of width $w$, oriented along the $x$-direction,
we have $\partial p/\partial x=-\eta\partial^{2}u_{x}/\partial y^{2}$.
\cite{LL_Fluid} Solving for $u_{x}$ with the boundary condition
(\ref{eq:General_boundary_condition}) we obtain
\begin{equation}
u_{x}=\frac{1}{8\eta}\left(w^{2}+4\zeta w-4y^{2}\right)\frac{\partial p}{\partial x},
\end{equation}
if $\partial p/\partial x$ is constant. The total heat current $I$
is proportional to the integral
\begin{equation}
I\propto\int_{-w/2}^{w/2}dy\,u_{x}=\frac{1}{2\eta}\left(\frac{w^{3}}{6}+\zeta w^{2}\right)\frac{\partial p}{\partial x}.\label{eq:Poiseuille_current}
\end{equation}
The second contribution stems from a finite slip velocity at the boundaries.
Only if $\zeta\ll w$, does the typical Poiseuille scaling $I\propto w^{3}$
(or $w^{d+1}$ for arbitrary dimensions) hold. Clearly, any hydrodynamic
effect, such as e.g. the Gurzhi effect (as the $I\propto w^{d+1}$
scaling is called in the context of electron flow), that depends on
stress created by momentum dissipation at the boundaries, is critically
influenced by $\zeta$. To illustrate the importance of boundary conditions,
expressed in terms of the slip length, we show in Fig. \ref{fig:Profiles_Slip_Length}
the flow profiles of a wire of thickness $w$ through which passes
a current $I$ for different slip lengths. 

In the context of electron hydrodynamics, the nature of the boundary
conditions is unclear. On the one hand, Poiseuille type flow, observed
in Refs. \cite{Moll2016,Gooth2017} supports at the least a very small
slip length if compared to the characteristic size of the system.
On the other hand, the absence of such flow in graphene was taken
as evidence for a no-stress boundary with a very large slip length
\cite{Bandurin}. 

In this paper we develop a kinetic theory to determine the slip length
$\zeta$ for Dirac and Fermi liquids in two limits. In the first limit
only a small fraction of tangetial momentum is transferred to the
wall in electron-wall collisions, which are assumed to be elastic.
This limit we call the nearly specular limit. In the opposite, the
diffuse, limit all tangential momentum is lost. The two limits correspond
to samples with almost smooth and strongly disordered edges. We find
that the slip lengths grow with decreasing temperatures. For graphene
at charge neutrality, in the nearly specular limit $\zeta\sim\ln^{6}\left(T_{\Lambda}/T\right)/T^{4}$,
whereas in the diffuse limit $\zeta\sim l_{ee}\sim\ln^{3}\left(T_{\Lambda}/T\right)/T$.
The logarithmic factor stems from a renormalization of the group velocity
of electrons caused by interaction effects \cite{Sheehy2007}. $T_{\Lambda}=\Lambda/k_{B}$,
where $\Lambda$ is an energy cut-off. For realistic parameters of
graphene, the slip lengths are larger than $1\,\upmu\textrm{m}$ below
$100\textrm{\,K}$, which leads us to the conclusion that for most
geometries used so far it is more appropriate to assume no-stress
than no-slip conditions at the boundaries. In addition we discuss
the slip length of three and two dimensional Fermi liquids, the latter
describing graphene at a finite chemical potential $\mu\gg k_{B}T$
and potentially the delafossite PdCoO$_{2}$ of Ref. \cite{Moll2016}.
We find that in three dimensions the slip length grows as $T^{-2}$
in both the diffuse scattering and the nearly specular scattering
limits, yet with a coefficient that depends on the nature of the boundary
scattering. For a two dimensional Fermi liquid the slip length behaves
as $T^{-2}/\ln\left(\epsilon_{F}/k_{B}T\right)$, due to the well
known logarithmic suppression of the quasiparticle liftime \cite{Hodges1971}.
Comparing Dirac and Fermi liquid, we find that in the diffuse scattering
limit, the slip lengths of graphene away from charge neutrality are
larger then those of charge neutral graphene below $100\textrm{\,K}$.
In the nearly specular limit, the slip length of charge neutral graphene
is larger than that of graphene away from charge neutrality at very
low temperatures, however, here $\zeta$ is very large for both systems.
If the momentum dissipation is due to edge roughness, we find that
graphene at a finite chemical potential $\mu$ is more susceptible
to the magnitude of the roughness, because here the electron wavelengths
- governed by the energy scale $\mu$ - are smaller than the thermal
electron wavelengths of charge neutral graphene. Summing up, we find
that the slip length for electronic flow can always be written in
the form 
\begin{equation}
\zeta=f\left(\kappa\right)l_{ee}
\end{equation}
with dimensionless ratio $\kappa=h^{2}h'^{d-1}/\lambda^{d+1}$ . $\kappa$
depends on the two length scales $h$ and $h'$ that characterize
the interface scattering (see Eq. (\ref{eq:Gaussian_roughness}) and
Fig. \ref{fig:The-disordered-edge}) and the electron wavelength $\lambda$,
respectively. The latter is strongly temperature dependent for graphene
at the neutrality point ($\lambda=\hbar v/(k_{{\rm B}}T)$), while
it corresponds to the Fermi wave length in the case of Fermi liquids
($\lambda=1/k_{{\rm F}}$). Here $v$ is the renormalized group velocity
of the electrons (see Eq. (\ref{eq:Graphene_Free_Hamiltonian}) and
the discussion below Eq. (\ref{eq:free_transform})). For the dimensionless
function $f\left(\kappa\right)$ we find $f\left(\kappa\ll1\right)=f_{0}/\kappa$,
while $f\left(\kappa\rightarrow\infty\right)\rightarrow f_{\infty}$.
We determine $f_{\infty}$ using the assumption of diffuse scattering.
The numerical values for the coefficients $f_{0}$ and $f_{\infty}$
depend sensitively on the electronic dispersion relation and dimensionality
of the system, but the overall behavior is found to be generic. We
find for Dirac systems at the neutrality point $f_{0}\approx0.008$
and $f_{\infty}\approx0.6$. For two-dimensional Fermi liquids holds
$f_{0}\approx1.1$ and $f_{\infty}\approx1.2$, while we obtain for
three-dimensional Fermi liquids $f_{0}\approx3$ and $f_{\infty}\approx0.5$.
For a boundary with intermediate scattering strength we expect $f\left(\kappa\right)$
to smoothly interpolate between the two limits, with a crossover for
$\kappa\sim{\cal O}\left(1\right)$.

\subsection*{Thermal currents in charge neutral graphene}

In a Galilei-invariant system, the drift velocity $\bm{u}$ is proportional
to the electric current:
\begin{equation}
\bm{j}=ne\bm{u},
\end{equation}
where $n$ is the particle density. This means that the hydrodynamic
flow of a Fermi liquid can be probed by measuring the electric current
$\bm{j}$. A key aspect of electron hydrodynamics in graphene at charge
neutrality is that here the heat current takes the place of mass or
charge current in conventional systems. The heat current is proportional
to the momentum density and therefore conserved in electron-electron
interactions. As a result the thermal conductivity at the neutrality
point is infinite, if the momentum is not dissipated by other mechanisms
such as impurities or boundary scattering \cite{M=0000FCller2008}.
This is a direct consequence of the linear dispersion of graphene.
The drift velocity $\bm{u}$ is connected to the heat current \cite{Briskot2015}
via 
\begin{equation}
\bm{j}_{E}=\frac{3n_{E}\bm{u}}{2+u^{2}/v^{2}}.
\end{equation}
Furthermore, at charge neutrality, no hydrodynamic flow $\bm{u}$
can be excited by applying an electric field because the same number
of hole-like excitations flows in one direction as electrons in the
other. A temperature difference, however, can be thermodynamically
related to a pressure difference. This qualitatively different behavior
of the thermal and electric AC conductivity is the reason for the
dramatic violation of the Wiedemann-Franz law observed in Ref. \cite{Crossno2016}.
Thus, a temperature gradient must be applied to a graphene sample
in order to excite a drift flow $\bm{u}$ \cite{Link2018}. To see
this, we use the differentials of the grand canonical potential and
the Gibbs-Duhem relation $\Omega=-pV$. One easily finds $\nabla p=s\nabla T+n\nabla\mu$,
where $n$ is the particle density and $s$ is the entropy density.
Both quantities, $s$ and $n$, are spatially uniform. The density
$n$ vanishes at charge neutrality (see Appendix A) and we are left
with
\begin{equation}
\nabla p=s\nabla T.\label{eq:pressure_temperature_relation}
\end{equation}
Then, in the linear and stationary regime, the Navier-Stokes equation
governing the incompressible hydrodynamic flow \cite{Fritz2008,Briskot2015}
reads
\begin{equation}
s\nabla T=\eta\Delta\bm{u},
\end{equation}
and the temperature gradient is playing the role of the external stress
that causes the flow. Such a situation is considered in section \ref{sec:Flow_Strip_Obstacle},
where we investigate the heat flow through an infinitely long strip
with an impenetrable circular obstacle. No-stress boundary conditions
are applied and viscous forces alone are seen to create a temperature
gradient. 

Finally, we add that the optical conductivity of Dirac systems \cite{Kashuba2008,Fritz2008}
can be considered to be a bulk signature of hydrodynamic behavior,
because the current relaxation mechanism is unrelated to momentum
conservation and therefore independent of boundary scattering. Furthermore,
the second-order nonlinear conductivity of a Dirac electron system
is expected to have unusual properties in the hydrodynamic regime
\cite{Sun2018}.

\section{Theory}

\subsection{Boundary conditions for the distribution function }

We now discuss how to obtain a boundary condition of the form of Eq.
(\ref{eq:General_boundary_condition}) from a kinetic theory. he techniaspects
of the determination of the viscosity are outlined in Appendix B.
Our program is to first determine the boundary conditions for the
underlying kinetic theory of the electron distribution function, developing
an understanding of how the kinetic distribution behaves at the boundary,
and then to connect to hydrodynamics. To be specific, we perform the
subsequent analysis for graphene at the neutrality point. Below we
also summerize the corresponding results for Fermi liquids.

The behavior of electrons in graphene is gouverned by the massless
Dirac Hamiltonian in two spatial dimensions
\begin{equation}
H_{0}=v_{0}\hbar\bm{k}\cdot\bm{\sigma}_{ab},\label{eq:Graphene_Free_Hamiltonian}
\end{equation}
and the Coulomb repulsion between electrons. $v_{0}$ is the bare
group velocity. $a$ and $b$ are sublattice (pseudospin) indices.
Here and hereafter we supress the valley degree of freedom. Together
with the spin degeneracy, we take it into account in the final results
as a global prefactor $N=4$. The non-interacting part of the graphene
Hamiltonian (\ref{eq:Graphene_Free_Hamiltonian}) is diagonalized
by a unitary transformation 
\begin{equation}
U_{\bm{k}}=\frac{1}{\sqrt{2}}\left[\begin{array}{cc}
o_{\bm{k}} & -o_{\bm{k}}\\
1 & 1
\end{array}\right],\label{eq:free_transform}
\end{equation}
where $o_{\bm{k}}=\left(k_{x}+ik_{y}\right)/\sqrt{k_{x}^{2}+k_{y}^{2}}$,
which results in a spectrum $\epsilon_{\lambda\bm{k}}=\lambda v_{0}\hbar k$.
$\lambda=\pm1$ is the band index. This spectrum exhibits a fourfold
spin-valley degeneracy. In the remainder of the text, whenever we
are concerned with graphene at the charge neutrality point, we use
the approach of \cite{Sheehy2007} in which interaction effects give
rise to a renormalization of the velocity $v_{0}\rightarrow v=v_{0}\left(1+\alpha\ln\left(\Lambda/k_{B}T\right)\right)$,
accompanied by the renormalization of the coupling constant $\alpha_{0}\rightarrow\alpha=e^{2}/(4\pi\epsilon\hbar v)$. 

\begin{figure}
\begin{centering}
\includegraphics[angle=270,scale=0.37]{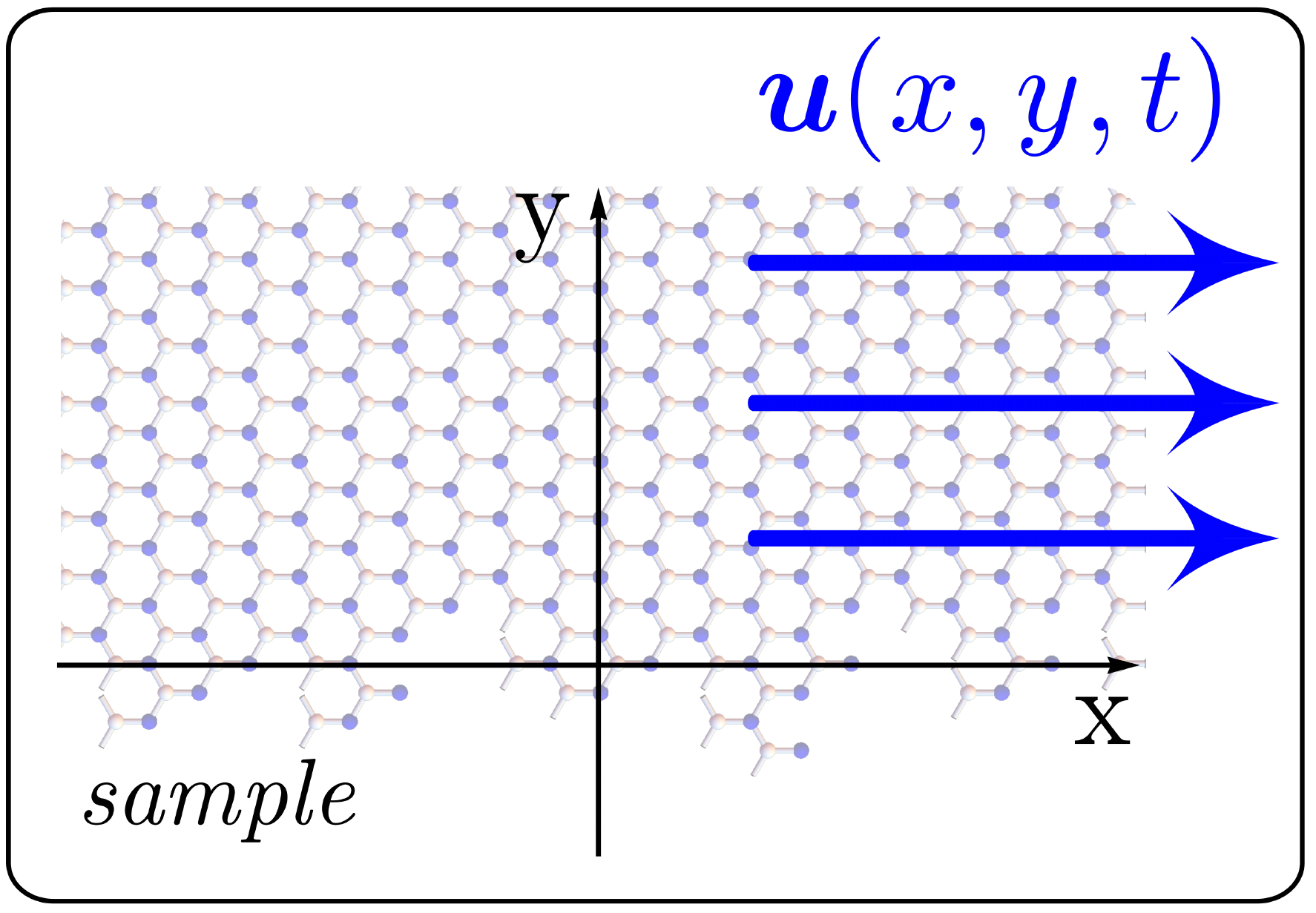}
\par\end{centering}

\caption{We consider the flow of an electron liquid through a graphene sample
that extends over the region $y>0$. The drift velocity is oriented
along the $x$-axis. The sample edge at $y=0$ is a source of momentum
dissipation.\label{fig:The_setting} }
\end{figure}

Consider a semi-infinite graphene sheet in the region $y>0$ with
an edge along the $x$-axis (Fig. \ref{fig:The_setting}). At a formal
level the kinetic equation for graphene electrons in the presence
of a boundary contains two collision terms: the electron-electron
collision term $C_{\textrm{e.e.}}\left[f\right]$ due to the Coulomb
interaction and the electron-edge collision term $C_{\textrm{edge}}\left[f\right]$.
In the absence of electric and magnetic fields the kinetic equation
takes the form
\begin{equation}
\left(\frac{\partial}{\partial t}-\bm{v}_{\lambda}\cdot\nabla\right)f_{\lambda,\bm{k}}\left(\bm{r}\right)=-C_{\textrm{e.e.}\lambda\bm{k}}\left[f\right]-C_{\textrm{edge}\lambda\bm{k}}\left[f\right],\label{eq:Full_Boltzmann}
\end{equation}
where $\bm{v}_{\lambda}=\lambda v\bm{k}/\left|k\right|$ is the group
velocity. The problem of solving Eq. (\ref{eq:Full_Boltzmann}) seems
rather challenging. It is, however, possible to reduce $C_{\textrm{edge}}\left[f\right]$
to a boundary condition for $f_{\lambda,\bm{k}}\left(x=0\right)$
\cite{OkulovUstinov1974}. This boundary condition relates the distribution
function of reflected electrons $f_{\bm{k}}^{>}$, which is defined
for $v_{\lambda}^{y}>0$, to that of the incident electrons $f_{\bm{k}}^{<}$,
defined for $v_{\lambda}^{y}<0$. Once $f_{\bm{k}}^{>}$ and $f_{\bm{k}}^{<}$
are found as solutions of the kinetic equation with the appropriate
boundary condition, the hydrodynamic boundary condition in the form
of Eq. (\ref{eq:General_boundary_condition}) follows from the fact
that the momentum current perpendicular to the impenetrable edge must
vanish at $y=0$: 
\begin{eqnarray}
0 & = & 2N\int_{<}\frac{d^{d}k}{\left(2\pi\right)}v_{+,\bm{k}}^{y}k_{x}f_{+,\bm{k}}^{<}\left(y=0\right)\nonumber \\
 &  & +2N\int_{>}\frac{d^{d}k}{\left(2\pi\right)}v_{+,\bm{k}}^{y}k_{x}f_{+,\bm{k}}^{>}\left(y=0\right).\label{eq:Mom_curr_average}
\end{eqnarray}
Here, the factor of two accounts for the particle and hole bands and
the factor of $N$ is due to the spin-valley degeneracy. The subscripts
$\gtrless$ denote that the integrals have to be taken over the regions
in momentum space where $v_{\lambda,\bm{k}}^{y}>0$, or $v_{\lambda,\bm{k}}^{y}<0$,
respectively. In order to derive (\ref{eq:General_boundary_condition})
we must take into account that the distribution functions $f_{\bm{k}}^{\lessgtr}$
depend on $\bm{u}$, as well as on its spatial derivatives. 

\subsubsection{Nearly specular limit}

Under the assumtion that the relevant momentum relaxation at the wall
stems from the irregular shape of the boundary, the boundary conditions
can be obtained from the scattering behavior of the electronic wave
function near a rough surface. Early phenomenological parametrizations
of the scattering behavior near such a surface go back to Maxwell
\cite{Maxwell1867} and Fuchs \cite{Fuchs1938}. Here, we follow a
more microscopic approach along the lines of Refs. \cite{Buchholtz1979,Falkovsky1983}. 

\begin{figure}
\begin{centering}
\includegraphics[angle=270,scale=0.44]{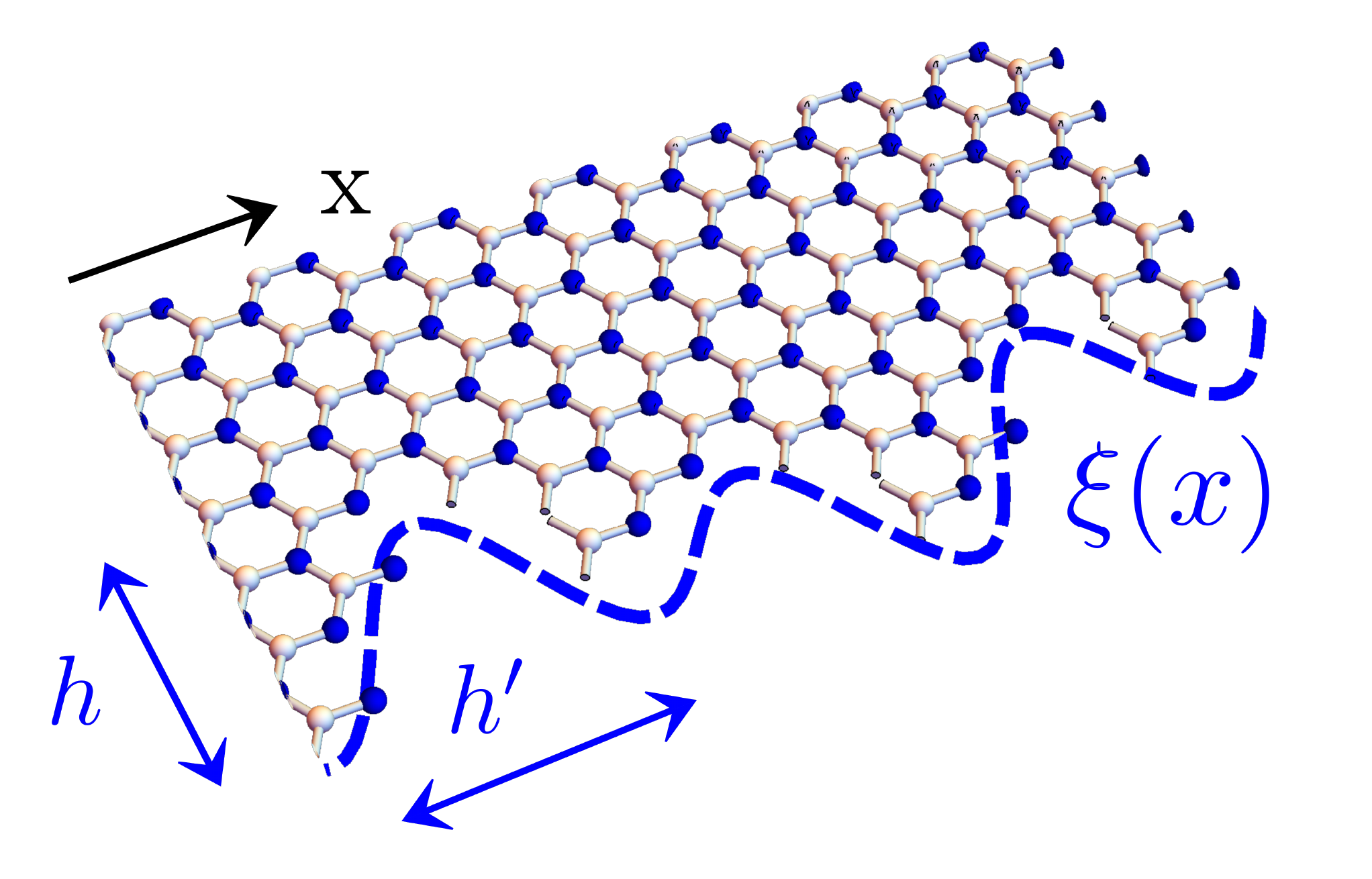}
\par\end{centering}

\caption{The disordered edge of a graphene sample is described by the function
$\xi\left(x\right)$ and characterized by its mean height $h$ and
correlation length $h'$. Both are much smaller than the thermal wavelenght
of the graphene electrons. On average, the edge can be described by
the distribution $\overline{\xi\left(x_{1}\right)\xi\left(0\right)}=h^{2}\exp\left(-\frac{x^{2}}{h'^{2}}\right)$.\label{fig:The-disordered-edge}}
\end{figure}

Let the rough surface be oriented along the $x$-axis and its shape
be given by the function $\xi\left(x\right)$ (see Fig. \ref{fig:The-disordered-edge}).
Before we can address the question of the boundary conditions for
the distribution function, we need to know the behavior of the Dirac
wave function, i.e. that of low energy excitations, at the sample
edge. It was shown in Ref. \cite{Akhmerov2008} within a tight-binding
approach that for almost any cut through the honeycomb lattice the
appropriate boundary condition is, quantitatively very similar to
that of the zig-zag edge. The only exception would be an armchair
edge that extends without disturbance over a large distance - something
wich is unprobable for a disordered edge and is excluded here. This
means that the wave function on one sublattice $a=1$ vanishes at
the boundary, whereas the wave function on the other sublattice $a=2$
remains undetermined, or vice versa. The valley degrees of freedom
are not mixed. To derive the boundary condition for the distribution
function $f_{\lambda,\bm{k}}$, it suffices to impose the condition
\begin{equation}
\psi_{a=1}\left(x,y=\xi\left(x\right)\right)=0.\label{eq:Wf_bound_cond}
\end{equation}
The deviations $\xi\left(x\right)$ must be larger than the interatomic
distances, but small compared to the typical wavelength of Dirac electrons
at low temperatures. The boundary condition (\ref{eq:Wf_bound_cond})
can then be expanded in $\xi\left(x\right)$:
\begin{equation}
\psi_{1}\left(x,0\right)+\xi\left(x\right)\frac{\partial\psi_{1}\left(x,y\right)}{\partial y}\Bigg|_{y=0}=0.\label{eq:Wf_bound_cond_taylor}
\end{equation}
We consider elastic scattering at the boundary only, therefore it
is usefull to introduce the projection of a wave function onto quasi-free
plane-wave states with a given energy $\epsilon$:
\begin{equation}
\psi_{1,\epsilon}\left(\bm{x}\right)=\sum_{\lambda}\int\frac{d^{2}k}{\left(2\pi\right)^{2}}\delta\left(\epsilon-\epsilon_{\lambda,k}\right)\gamma_{\bm{k},\lambda}U_{1,\lambda}\left(\bm{k}\right)e^{i\bm{k}\cdot\bm{r}}.
\end{equation}
Here, $\epsilon_{\lambda,k}=\lambda v\hbar k$ is the electron dispersion
and $U_{a,\lambda}\left(\bm{k}\right)$ transforms the wavefunctions
$\gamma_{\bm{k},\lambda}$ from the band basis into the sublattice
basis. $\psi_{\bm{k},a}=\sum_{\lambda}\gamma_{\bm{k},\lambda}U_{a,\lambda}\left(\bm{k}\right)$
is the Bloch function projected onto the sublattice. Inserting $\psi_{1,\epsilon}\left(\bm{x}\right)$
into (\ref{eq:Wf_bound_cond_taylor}), carrying out the $k_{x}$ integration,
and performing a Fourier transform, we obtain to second order in $\xi\left(x\right)$
a relation between the wavefunctions $\gamma_{k_{x},\left|k_{y}\right|,\lambda}$
and $\gamma_{k_{x},-\left|k_{y}\right|,\lambda}$ on the boundary.
This relation holds for $\lambda=\pm1$ separately, because of the
elasticity of the scattering process. Then, an average over the edge
shapes $\overline{\xi\left(x_{1}\right)\xi\left(x_{2}\right)}$ is
taken, so that translation invariance along the edge is restored:
\begin{equation}
\overline{\xi\left(k_{1}\right)\xi\left(k_{2}\right)}=2\pi\delta\left(k_{1}+k_{2}\right)W\left(k_{1}\right).
\end{equation}
Thus, $W\left(k\right)$ describes the correlation of the surface
roughness. The squared moduli of the wavefunctions are directly related
to the kinetic distribution function on the boundary:
\begin{equation}
f_{\bm{k},\lambda}=\left(2\pi v_{k_{x},\lambda}^{y}\right)^{-1}\overline{\left|\gamma_{k_{x},k_{y}\left(\epsilon,k_{x}\right),\lambda}\right|^{2}}.
\end{equation}
The prefactor $\left(2\pi v_{k_{x},\lambda}^{y}\right)^{-1}$ stems
from a variable change $k_{x},\epsilon\rightarrow\bm{k}$. In this
fashion we arrive at the boundary condition 
\begin{eqnarray}
f^{>}\left(k_{x},k_{y}\right) & = & f^{<}\left(k_{x},-k_{y}\right)\label{eq:Falktovsky_conditions}\\
 &  & -4f^{<}\left(k_{x},-k_{y}\right)k_{y}\int\frac{dk'_{x}}{2\pi}k'_{y}W\left(k_{x}-k'{}_{x}\right)\nonumber \\
 &  & +4k_{y}\int\frac{dk'_{x}}{2\pi}k'_{y}W\left(k_{x}-k'{}_{x}\right)f^{<}\left(k'_{x},-k'_{y}\right),\nonumber 
\end{eqnarray}
where $\gtrless$ stands for the sign of the velocity component in
the $y$-direction. Except for the matrix elements $U_{a,\lambda}\left(\bm{k}\right)$,
which ultimately cancel out, and the fact that two bands $\lambda=\pm1$
have to be kept track of, the calculation is completely analogous
to the one presented by Falkovsky in \cite{Falkovsky1983}. The domain
of integration in (\ref{eq:Falktovsky_conditions}) ranges from $k'_{x}=-\epsilon/v$
to $k'_{x}=\epsilon/v$ where $k_{y}=\sqrt{\left(\epsilon/v\right)^{2}-k_{x}^{2}}$.
Interchanging the sublattice index in Eq. (\ref{eq:Wf_bound_cond})
does not alter the result of Eq. (\ref{eq:Falktovsky_conditions}).

We assume that the edge correlation function $\overline{\xi\left(x_{1}\right)\xi\left(x_{2}\right)}$
is given by a Gaussian distribution
\begin{equation}
\overline{\xi\left(x_{1}\right)\xi\left(0\right)}=h^{2}e^{-\frac{x^{2}}{h'^{2}}},\label{eq:Gaussian_roughness}
\end{equation}
where $h$ is the typical amplitude of $\xi(x)$ and $h'$ is their
correlation length. We then have
\begin{equation}
W\left(k_{x}\right)=\sqrt{\pi}h^{2}h'e^{-\frac{1}{4}k_{x}^{2}h'^{2}}.
\end{equation}
In graphene at charge neutrality the characteristic energy of excitations
is $\epsilon_{T}\sim k_{B}T$. If the lengths $h$, $h'$ are of the
order of a few interatomic distances, we can safely assume for the
thermal wavelength $\lambda_{T}=v\hbar/\epsilon_{T}$ that 
\begin{equation}
h'\ll\lambda_{T},
\end{equation}
and therefore that $W\left(k_{x}\right)$ is a flat function: 
\begin{equation}
W\left(k_{x}\right)\approx\sqrt{\pi}h^{2}h'.
\end{equation}
The presence of the small parameter $h^{2}h'/\lambda_{T}^{3}$ is
the reason, why our analysis of the slip length in the nearly specular
limit is well controlled. The boundary condition (\ref{eq:Falktovsky_conditions})
does conserve the number of particles and has essentially the form
of the boundary condition proposed by Fuchs \cite{Fuchs1938}, where
\begin{equation}
p\left(\bm{k}\right)=1-4k_{y}\int\frac{dk'_{x}}{2\pi}k'_{y}W\left(k_{x}-k'{}_{x}\right)
\end{equation}
 takes the role of a specularity parameter which depends on the angle
of incidence.

\subsubsection{Diffuse limit}

An alternative boundary condition is valid in the limit of totaly
diffuse boundary scattering. Here, it is sufficient to assume that
the distribution of electrons departing from the wall does not depend
on the particle directions, i.e. 
\begin{equation}
f_{\lambda,\bm{k}}^{>}=f_{\lambda}^{>}\left(\left|\bm{k}\right|\right).\label{eq:Diffuse_conditions}
\end{equation}
Clearly, in such a case all tangential momentum is lost in an electron-wall
collision. The diffuse limit is appropriate, if the sample edge is
very rough and one makes no assumption on the elasticity of the scattering
processes.

\subsection{Kinetic equation at a boundary\label{sub:Kinetic-equation-at-boundary}}

Next, we determine the electron flow behavior that characterizes the
transfer of momentum near the surface within a kinetic theory. Generally,
within the Chapman-Enskog approach \cite{Chapman1970,Cercignani1988},
the bulk kinetic distribution function has the form
\begin{equation}
f\big|_{b}=f^{l.e.}-\frac{\partial f^{0}}{\partial\epsilon}\Psi,\label{eq:General_Chapman}
\end{equation}
where $f^{l.e.}$ is the local equilibrium distribution function,
which is found by setting the collision integral to zero, and $f^{0}$
the distribution function for the global equilibrium. For graphene
electrons
\begin{equation}
f_{\lambda,\bm{k}}^{l.e.}=\frac{1}{e^{\beta\left(\epsilon_{\lambda,\bm{k}}-\bm{u}\left(\bm{r},t\right)\cdot\bm{k}-\mu\right)}+1}
\end{equation}
and $f^{0}$ is the Fermi-Dirac distribution. The inverse temperature
$\beta=1/(k_{B}T)$ is not to be confused with the coordinate index
$\beta$. $\Psi$ is a non-equilibrium contribution describing the
response to shear and other forces. The response to shear forces is
characterized by the viscosity $\eta$ of a system, defined via the
relation
\begin{equation}
\tau_{\alpha\beta}=-\eta\frac{\partial u_{\alpha}}{\partial x_{\beta}}
\end{equation}
between the stress (momentum current) tensor $\tau_{\alpha\beta}=N\hbar v\sum_{\lambda}\int\frac{d^{2}k}{\left(2\pi\right)^{2}}\left(\lambda k_{\alpha}k_{\beta}/k\right)f_{\mathbf{k}\lambda}$
and the gradient of the drift velocity. In the absence of a wall,
the kinetic distribution function for graphene at charge neutrality
and due to electron-electron Coulomb interaction was calculated in
\cite{Mueller2009} (the main points are summerized in Appendix B).
In the presence of shear forces only, the bulk distribution to leading
order in the fine structure constant $\alpha=e^{2}/\left(\epsilon\hbar v\right)$
(not to be confused with the coordinate index $\alpha$) and the drift
velocity $\bm{u}$ is given by 
\begin{align}
f_{\lambda,\bm{k}}\big|_{b} & =f_{\lambda,\bm{k}}^{l.e.}+\beta f_{\lambda,\bm{k}}^{0}\left(1-f_{\lambda,\bm{k}}^{0}\right)\Psi,\label{eq:Bulk_distribution_graphene}
\end{align}
where
\begin{eqnarray}
\Psi & = & \frac{\lambda}{2\sqrt{2}}\left(C_{0}+C_{1}\beta vk\right)I_{\alpha\beta}X_{\alpha\beta}\nonumber \\
X_{\alpha\beta} & = & \frac{\partial u_{\alpha}}{\partial x_{\beta}}+\frac{\partial u_{\beta}}{\partial x_{\alpha}}-\delta_{\alpha\beta}\nabla\cdot\bm{u}\nonumber \\
I_{\alpha\beta} & = & \sqrt{2}\left(\frac{k_{\alpha}k_{\beta}}{k^{2}}-\frac{1}{2}\delta_{\alpha\beta}\right).\label{eq:Chapman_ansatz}
\end{eqnarray}
Here, $C_{0}$ and $C_{1}$ are dimensionless numerical coefficients
- the $\psi_{0}$ and $\psi_{1}$ of Appendix B - that are found by
solving the kinetic equation \cite{Mueller2009} (see Appendix B).
$C_{0}$ and $C_{1}$ correspond to the zero modes of the collinear
part of the collision integral and are dominant at leading order in
the fine structure constant $\alpha$. The expression (\ref{eq:Bulk_distribution_graphene})
can be used to determine the viscosity of graphene electrons
\begin{equation}
\eta=N\frac{\left(\pi^{3}C_{0}+27\zeta(3)\pi C_{1}\right)}{48\pi^{2}\beta^{2}v^{2}\hbar}\approx\frac{0.449N}{4\alpha^{2}v^{2}\hbar}\left(k_{B}T\right)^{2},
\end{equation}
with $N=4$ being the spin valley degeneracy.

In the presence of the sample edge, we expect corrections of the order
$h^{2}h'$ to the bulk distribution function stemming from the edge,
therefore we make for the distribution function $f_{\lambda,\bm{k}}^{<}\left(y=0\right)$
of particles impinging on the edge the ansatz 
\begin{equation}
f_{\lambda,\bm{k}}^{<}\left(y=0\right)=f_{\lambda,\bm{k}}\big|_{b}+\mathcal{O}\left(h^{2}h'k_{T}^{3}\right)A\left(I_{\alpha\beta}X_{\alpha\beta}\right),\label{eq:Distiribution_Ansatz}
\end{equation}
where $A\left(I_{\alpha\beta}X_{\alpha\beta}\right)$ is some function
of gradients of the drift velocity $\bm{u}$ and momenta $k^{i}$.
As we will show later, this correction contributes to the slip-length
only to second order in $h^{2}h'$ and we can ignore the contribution
$A\left(I_{\alpha\beta}X_{\alpha\beta}\right)$. In other words, one
can safely assume that the distribution function of the electrons
that move towards the sample edge is governed by the bulk distribution
function. Thus, the loss of tangential momentum is described by the
boundary condition (\ref{eq:Falktovsky_conditions}) and we do not
need to make any assumtions on the influence of the boundary on momentum
currents. Inserting (\ref{eq:Distiribution_Ansatz}) into (\ref{eq:Falktovsky_conditions})
we obtain an expression for $f_{\lambda,\bm{k}}^{>}\left(y=0\right)$.
In this way we know the distribution function at the edge $f_{\lambda,\bm{k}}\left(y=0\right)$.

\subsection{The slip length \label{sub:The-slip-length}}

\subsubsection{The nearly specular limit}

Knowing the functions $f_{\lambda,\bm{k}}^{<}\left(y=0\right)$ and
$f_{\lambda,\bm{k}}^{>}\left(y=0\right)$ as a function of $u_{x}\left(y=0\right)$
and $\left.\partial_{y}u_{x}\right|_{y=0}$, we find the hydrodynamic
boundary condition with the help of Eq. (\ref{eq:Mom_curr_average}).
It is also possible to obtain a boundary condition of the form (\ref{eq:General_boundary_condition})
by avereging over the momentum: 
\begin{eqnarray}
\frac{3n_{E}}{2v^{2}}u_{x} & = & 2N\int_{<}\frac{d^{d}k}{\left(2\pi\right)}k_{x}f_{+,\bm{k}}^{<}\left(y=0\right)\nonumber \\
 &  & +2N\int_{>}\frac{d^{d}k}{\left(2\pi\right)}k_{x}f_{+,\bm{k}}^{>}\left(y=0\right).\label{eq:Mom_average}
\end{eqnarray}
Note, that the drift velocity is related to the momentum density $\bm{g}_{\bm{k}}=\sum_{\lambda}\int d^{2}k'/\left(2\pi\right)^{2}\left(\hbar\bm{k}\right)f_{\lambda,k}$
via $\left(3n_{E}/2v\right)\bm{u}\approx\bm{g}_{\bm{k}}$, where $n_{E}$
is the energy density \cite{Briskot2015}. In the nearly specular
limit, the two approaches (\ref{eq:Mom_average}) and (\ref{eq:Mom_curr_average})
give the same result. In the diffuse scattering limit the second equation
(\ref{eq:Mom_curr_average}) will give the better result, because
the additional factor $\sin\left(\varphi\right)$ in the integrands
- due to $v_{+,\bm{k}}^{y}$ - gives more weight to contributions
from particles with an incidence angle near $\pi/2$. These particles
are least influenced by the Knudsen boundary layer - a one mean free
path broad region along the sample edge where the distribution function
significantly deviates from its bulk values. Therefore our assumption
that the loss of tangential momentum is determined directly at the
wall, by the boundary condition (\ref{eq:Falktovsky_conditions}),
is appropriate here. 

Performing the average, we see that only those parts of $f_{\lambda,\bm{k}}\left(y=0\right)$
contribute, which are proportional to $\cos\left(\varphi\right)$.
Therefore, for a flat $W\left(k_{x}\right)$, the last right hand
side term of (\ref{eq:Falktovsky_conditions}) does not contribute
to the momentum current average (\ref{eq:Mom_curr_average}). After
performing the integrations we have from (\ref{eq:Mom_curr_average})
\begin{equation}
0=-\left(\frac{h^{2}h'}{\lambda_{T}^{3}}\right)Au_{x}+\eta\frac{\partial u_{x}}{\partial y}-\left(\frac{h^{2}h'}{\lambda_{T}^{3}}\right)B\frac{\partial u_{x}}{\partial y},
\end{equation}
where we have defined the thermal wavelength $\lambda_{T}=\beta v\hbar$
and $A=\frac{N31\pi^{11/2}}{672\beta^{3}v^{3}\hbar^{2}}.$ It is again
clear, that $h^{2}h'/\lambda_{T}^{3}$ plays the role of a small dimensionless
parameter. Physically, the presence of the small parameter $h^{2}h'/\lambda_{T}^{3}$
shows that the edge behaves as if it was smooth, if its roughness
is on average much smaller than the typical wavelength of scattered
electrons. Solving for $u_{x}$, we write the above equation in the
form of (\ref{eq:General_boundary_condition}). To leading order in
$h^{2}h'/\lambda_{T}^{3}$ the slip length is given by
\begin{equation}
\zeta=\left(\frac{\lambda_{T}^{3}}{h^{2}h'}\right)\chi\approx0.008\left(\frac{\lambda_{T}^{3}}{h^{2}h'}\right)l_{ee},
\end{equation}
where $\chi=\frac{672\beta^{3}v^{3}\hbar^{2}}{N31\pi^{11/2}}\eta$.
We used $l_{ee}=v\hbar/\left(\kappa_{1}\alpha^{2}k_{B}T\right)$ for
the mean free path due to electron scattering with numerical coefficient
$\kappa_{1}=1.950$ (see Appendix B).

\subsubsection{The diffuse limit}

The overall procedure to find the boundary condition in the diffuse
limit is analogous to the nearly specular case. Due to the condition
(\ref{eq:Diffuse_conditions}) only impinging particles with a negative
velocity contribute to the average over the momentum current. In distinction
to the nearly specular case however, we do not have a small parameter
and therefore assume that the incident electron's behavior is described
by the bulk distribution function up to right at the edge. In the
theory of classical gases, this assumption leads to the famous Maxwell
boundary condition \cite{Maxwell1867} for rarified gas flow (see
also \cite{Kramers1949,Kennard1938}). The momentum current averaged
over the distribution function given in Eq. (\ref{eq:Diffuse_conditions})
yields
\begin{align}
\zeta=\frac{\pi^{2}\beta^{3}v^{3}\hbar^{2}}{3N\zeta(3)}\eta & \approx0.6l_{ee}.\label{eq:Diffuse_slip_length}
\end{align}
Again, we used the electron mean free path $l_{ee}=v\hbar/\left(\kappa_{1}\alpha^{2}k_{B}T\right)$.

\subsubsection{Discussion}

\begin{figure}
\centering{}\includegraphics[scale=0.45]{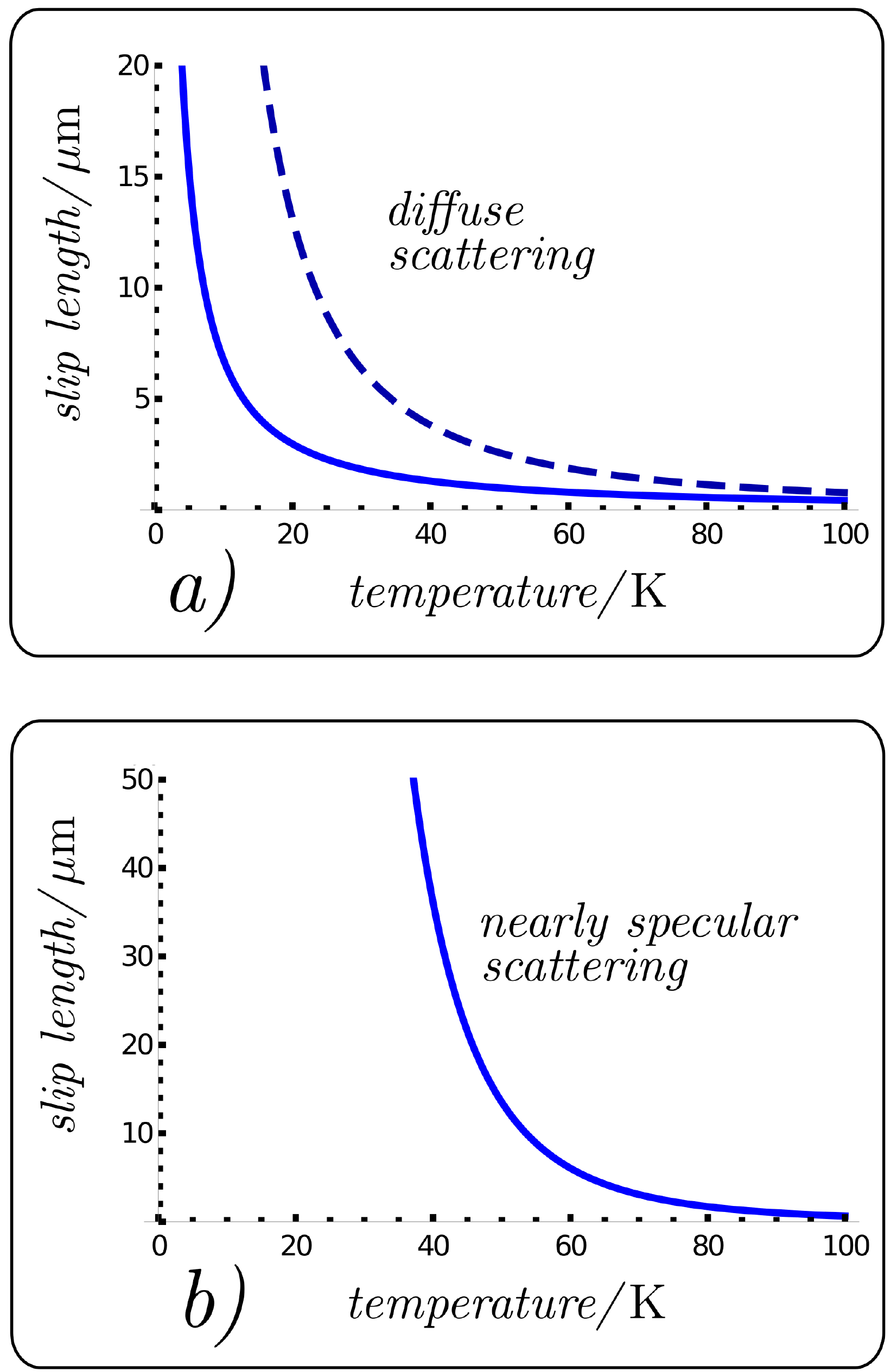}\caption{The figure shows the temperature dependence of the slip length $\zeta$
for a) diffuse scattering at the boundary for graphene at charge neutrality
(continuous line) and at finite chemical potential (dashed line) and
b) scattering at a microscopically rough edge with a typical roughness
of $h=h'=250\,\textrm{\AA}$ for graphene at charge neutrality. With
decreasing temperature, $\zeta$ grows as $T^{-4}\ln^{6}\left(T_{\Lambda}/T\right)$
in the nearly specular and as $T^{-1}\ln\left(T_{\Lambda}/T\right)$
in the diffuse limits at charge neutrality. Away from charge neutrality
the slip length behaves as $\zeta\sim T^{-2}/\ln\left(\epsilon_{F}/T\right)$
in both limits, but is parametrically larger in the nearly specular
case.\label{fig:Slip_length_plots}}
\end{figure}

In the diffuse limit, as well as in the nearly specular limit the
slip length $\zeta$ approaches infinity as $T\rightarrow0$. While
in the diffusive limit $\zeta\sim T^{-1}\ln^{3}\left(T_{\Lambda}/T\right)$
with $T_{\Lambda}=\Lambda/k_{B}$, in the nearly specular limit holds
$\zeta\sim T^{-4}\ln^{6}\left(T_{\Lambda}/T\right)$, showing clearly
that the mechanism of scattering on a rough boundary is ineffective
for electrons with large wavelength. The slip lengths as functions
of temperature are shown in Fig. \ref{fig:Slip_length_plots}.
In the renormalization of the velocity $v_{0}\rightarrow v=v_{0}\left(1+\alpha\ln\left(\Lambda/k_{B}T\right)\right)$
and the coupling constant $\alpha\rightarrow\alpha_{r}=e^{2}/(4\pi\epsilon\hbar v)$
we used a cut-off of $\Lambda\sim1\,\textrm{eV}$. Also, we assumed
a permittivity $\varepsilon=5\varepsilon_{0}$. Our small parameter
for the nearly specular limit $h/\lambda_{T}$ remains, below $100\,\textrm{K}$,
small up to an $h\approx250\,\textrm{\AA}$ (where it is $\approx1/5$
at $100\,\textrm{K}$).

In the diffusive limit, for the same parameter values as above, $\zeta$
ranges from $100\,\upmu\textrm{m}$ at $1\text{K}$ to $0.4\,\upmu\textrm{m}$
at $100\textrm{K}$. In the nearly specular limit, for a small roughness
of the order of $h=h'=10\,\textrm{\AA}$, $\zeta$ is comparable to
the length of the Trans-Siberian Railway at $T=1\,\mbox{K}$ and ranges
to $1\,\textrm{mm}$ at $T=100\,\text{K}$. For a fairly rough edge
of $h=h'=250\,\textrm{\AA}$ we have at $1\,\text{K}$ $\zeta=3.5\,\textrm{km}$
and at $T=100\,\text{K}$ the slip length $\zeta\approx0.6\,\upmu\textrm{m}$
approaches the diffuse limit. Such large values for $\zeta$ imply
that one can effectively use no-stress boundary conditions. 

We finally note, that the specularity of different kinds of edges
of different materials is well studied \cite{Tsoi1999}. For oxygen-plasma-etched
graphene specifically, values of 0.2 to 0.5 were reported for the
specularity parameter $q$ (which gives the probability that a single
scattering event at the edge is specular) \cite{Taychatanapat2013}.
Therefore, under these particular conditions the slip lengths are
expected to lie somewhere between the nearly specular and the diffuse
scattering limits.

\subsection{Fermi liquids and graphene away from charge neutrality}

Our derivation of boundary conditions for the hydrodynamic flow of
a Dirac liquid can also be applied to the Fermi liquids, which includes
graphene far away from charge neutrality. Let us again assume that
the $y$-axis is orthogonal to the boundary of the sample, that the
Fermi liquid is contained in the region $y>0$, and that the flow
is tangential to the boundary. 

Following \cite{Abrikosov1959}, we write the distribution function
of quasiparticles as
\begin{equation}
f=f_{0}\left(\epsilon\right)-\frac{\partial f_{0}}{\partial\epsilon_{0}}\Psi.
\end{equation}
Here, $\epsilon$ is the full quasiparticle energy which itself depends
on the occupation numbers and $\epsilon_{0}=v_{F}\hbar\left(k-k_{F}\right)+\epsilon_{F}$.
The funtion $\Psi$ describes the response to gradients of the drift
velocity $\bm{u}$. In the considered geometry, it can be parametrized
as
\begin{equation}
\Psi=+q\left(p\right)p_{x}\frac{\partial\epsilon}{\partial p_{y}}\frac{\partial u_{x}}{\partial y}.
\end{equation}
The stress tensor is given by 
\begin{equation}
\tau_{xy}=\int\frac{d^{d}k}{\left(2\pi\right)^{d}}\left(p_{x}v_{F}^{y}\right)\frac{\partial f_{0}}{\partial\epsilon_{0}}\Psi.
\end{equation}
Comparing to the relation $\tau_{xy}=-\eta\partial u_{x}/\partial y$,
we find for the viscosity the expression
\begin{equation}
\eta=-\int\frac{d^{d}k}{\left(2\pi\right)^{d}}\left(p_{x}v_{F}^{y}\right)^{2}q\left(p\right)\frac{\partial f_{0}}{\partial\epsilon_{0}}.
\end{equation}
For $d=3$ we then have
\begin{equation}
\eta=\frac{2}{15}v_{F}^{2}\rho_{F}\epsilon_{F}m^{*}\tau
\end{equation}
and for $d=2$
\begin{equation}
\eta=\frac{1}{4}v_{F}^{2}\rho_{F}\epsilon_{F}m^{*}\tau,
\end{equation}
where $\rho_{F}$ is the density of states at the Fermi surface and
\begin{equation}
\tau=\int_{-\infty}^{\infty}dx\frac{q\left(x\right)}{\left(2\cosh\left(\frac{x}{2}\right)\right)^{2}},
\end{equation}
with the dimensionless integration variable $x=\epsilon-\epsilon_{F}/\left(k_{B}T\right)$.
The quantity $q\left(x\right)$ must be found by solving the linearized
kinetic equation. 

For $d=3$ it was shown in \cite{Abrikosov1959} that $q\left(x\right)$
can be assumed constant. This yields that the leading temperature
dependence at $\epsilon_{F}\gg k_{B}T$ is $\eta\propto q\propto T^{-2}$.
where $v_{F}$ is the Fermi velocity. The detailed expressions for
the kinetic distribution function and viscosity can be found in \cite{Abrikosov1959}.
Compared to the case of graphene, the boundary condition (\ref{eq:Falktovsky_conditions})
holds as it is, except for the fact that the integrations have to
be performed over a two-dimensional surface and only the $\lambda=+1$
part is relevant. Furthermore $N=2$ due to the spin degeneracy.

In the nearly specular limit the role of the thermal wavelength $\lambda_{T}$
is played by the Fermi-wavevector $k_{F}$, as it determines the characteristic
wavelength of the exitations. The slip length as derived from (\ref{eq:Mom_curr_average})
to leading order in the parameter $1/h^{2}h'^{2}k_{F}^{4}$ and to
leading order in temperature is
\begin{equation}
\zeta=\left(\frac{1}{h^{2}h'^{2}k_{F}^{4}}\right)\chi\approx\left(\frac{1}{h^{2}h'^{2}k_{F}^{4}}\right)3l_{ee},
\end{equation}
where $\chi=\frac{45\pi^{2}}{k_{F}^{4}\hbar}\eta$. In the diffuse
scattering limit, we find
\begin{equation}
\zeta=\frac{8\pi^{2}}{k_{F}^{4}\hbar}\eta\approx0.5l_{ee}.\label{eq:Fermi_liquid_diffuse_boundary_cond}
\end{equation}
We used $l_{ee}=v_{F}\tau$ and $\rho_{F}=m^{*}k_{F}/\left(\hbar\pi\right)^{2}$.
The temperature dependence of the slip length in the diffuse scattering
limit as well as in the nearly specular limit is $\zeta\sim1/T^{2}$. 

Poiseuille type flow was observed in the delafossite PdCoO$_{2}$
\cite{Moll2016}. The same publication reports a viscosity of up to
$6\cdot10^{-3}\,\textrm{kg}/(\textrm{ms})$. With the help of Eq.
(\ref{eq:Fermi_liquid_diffuse_boundary_cond}) we find a slip length
of $\zeta=0.45\,\upmu\textrm{m}$. This length is indeed small compared
to the sample widths of up to $60\,\upmu\textrm{m}$, meaning that
the slip velocity at the boundaries is negligable, which, again, is
fully consistent with the observed Poiseuille behavior. A comparable
value for the electron viscosity and Poiseuille type flow in the Weyl
material WP$_{2}$ was reported in \cite{Gooth2017}. The sample widths
exeeded the slip length as given by (\ref{eq:Fermi_liquid_diffuse_boundary_cond})
and the observations were consistent with our theory. A typical Fermi
liquid, however, is expected to have a higher viscosity and a larger
slip length.

For $d=2$ we can crudely estimate the time $\tau$ by the quasiparticle
lifetime $\tau_{qp}$, which is known to be logarithmically suppresed
at low temperatures compared to the $d=3$ result \cite{Hodges1971}.
From the Refs. \cite{Reizer1997,Zheng1996,Menashe1996} we expect
\begin{equation}
\tau_{qp}=A\frac{\epsilon_{F}\hbar}{\left(k_{B}T\right)^{2}}\frac{1}{\ln\left(\frac{\epsilon_{F}}{k_{B}T}\right)},\label{eq:Estimate_lifetime}
\end{equation}
where $A$ is a coefficient of the order of unity. From (\ref{eq:Mom_curr_average})
we obtain in the nearly specular limit
\begin{equation}
\zeta=\left(\frac{\hbar^{3}v_{F}^{3}/\epsilon_{F}^{3}}{h^{2}h'}\right)\chi\approx\left(\frac{1}{h^{2}h'k_{F}^{3}}\right)1.1l_{ee},
\end{equation}
with $\chi=\frac{2}{\sqrt{\pi}}v_{F}\tau$, and in the diffuse limit
$\zeta=\frac{3\pi}{8}v_{F}\tau$, so that
\begin{equation}
\zeta\approx1.2l_{ee}.\label{eq:Doped_diffusive_sliplength}
\end{equation}
These results also apply to graphene at finite chemical potential.
The viscosity of graphene can then be written $\eta=\frac{1}{8}\rho_{F}^{G}\mu^{2}\tau$.
Using the quasiparticle lifetime (\ref{eq:Estimate_lifetime}) with
$A=1$ to estimate $\tau$, the slip lengths for graphene at $\mu\gg k_{B}T$
in the diffusive limit are larger than for graphene at charge neutrality.
For $\mu=0.25\textrm{eV}$, they range from $0.8\,\upmu\textrm{m}$
at $100\,\textrm{K}$ to $3\textrm{mm}$ at $1\,\textrm{K}$ (see
Fig \ref{fig:Slip_length_plots}). The reason for the larger slip
lengths is the $1/\left(T^{2}\ln\left(\epsilon_{F}/k_{B}T\right)\right)$
temperature dependence of $\tau$. In the nearly specular case, the
small parameter $\hbar^{3}v_{F}^{3}/\left(h^{2}h'\epsilon_{F}^{3}\right)=1/(h^{2}h'k_{F}^{3})$
does not depend on the thermal wavelength $\lambda_{T}$, but instead
on the wavelength at the Fermi surface. Since the edge roughness $h$
has to be compared to $k_{F}^{-1}\ll\lambda_{T}$, the diffusive limit
- giving the minimal $\zeta$, since all tangential momentum is lost
- can be saturated for much smaller $h$ than at charge neutrality.
Still, $\zeta$ as given by (\ref{eq:Doped_diffusive_sliplength})
is large enough to justify no-stress conditions for most geometries. 

Our result explains the findings of \cite{Bandurin}, where in graphene
samples with widths up to $4\,\upmu\textrm{m}$ no Gurzhi effect was
observed up to $100\,\textrm{K}$ and strengthen the author's conjecture
that the small deviation of the resistivity curves at about $100\,\textrm{K}$
indeed stems from a small hydrodynamic contribution due to the Gurzhi
effect (supplement to \cite{Bandurin}). The reason is that at about
$100\,\textrm{K}$ the slip length drops below $1\,\upmu\textrm{m}$
and becomes smaller than the sample width.

Let us finally note that the diffusive boundary condition gives the
smallest possible slip length and is a ``worst case scenario'' in
the sense that all tangential momentum is lost; be it due to elastic
or inelastic scattering. While the nearly specular scattering limit
deals with the opposite scenario and elastic scattering only, one
could imagine that considering larger and larger roughnesses $h$,
the slip length would saturate at some value $\zeta_{min}$, which
is close to the slip length of the diffusive scattering limit.

\section{Comparison to known results}

Most calculations of slip in quantum fluids \cite{Levinson1977,HJensen1980,Einzel1984,Einzel1990,Einzel1997}
model interactions by a momentum conserving relaxation time ansatz,
similar to the one used in the Bhatnagar-Gross-Krook equation \cite{Bhatnagar1954}.
The collision integral is replaced by the term $-g/\tau$, where $g$
is the deviation of the distribution function not from the global,
but from the local equilibrium: 
\begin{equation}
g=f-f^{l.e.}.\label{eq:d_def}
\end{equation}
Within this approach, it was shown in \cite{HJensen1980} for a diffusely
scattering boundary that the slip length determined in analogy to
the Maxwell slip length (see \cite{Maxwell1867,Kramers1949,Kennard1938}),
i. e. by assuming the validity of the bulk distribution function for
ingoing particles up to the boundary, gives a lower bound on the slip
length $\zeta$ as calculated within the Bhatnagar-Gross-Krook-like
approach. For completeness, we want to summerize the logic of \cite{HJensen1980}
briefly and discuss how it relates to our results. 

The analysis applies to two or three dimsensions and to arbitrary
dispersion relations. Therefore we will not specify dimensionality
and dispersion until the end. Let the $y$-axis be orthogonal to the
sample boundary and the quantum fluid be contained in the volume $y>0$.
The kinetic equation becomes
\begin{equation}
v_{y}\frac{\partial g}{\partial y}-v_{y}p_{x}\frac{\partial f^{0}}{\partial\epsilon}\frac{\partial u_{x}}{\partial y}=-\frac{g}{\tau}\label{eq:BGK_kin_eq}
\end{equation}
and is a first order differential equation for $g$ which is easily
solved as soon as the appropriate boundary conditions are formulated.
The idea is to describe the physics of the Knudsen layer right at
the boundary, where the system is not in local equilibrium, but significantly
influenced by the scattering of particles at the sample edge. At $y\rightarrow\infty$
the Knudsen layer ends and the system enters the hydrodynamic regime
described by the Navier-Stokes equations. Therefor the gradient of
the drift velocity $u'=\partial u_{x}/\partial y$ approaches a finite
value $u'\left(\infty\right)$ and it holds
\begin{equation}
g\left(\infty\right)=\tau v_{y}p_{x}\frac{\partial f^{0}}{\partial\epsilon}u'\left(\infty\right).
\end{equation}
\begin{figure}
\begin{centering}
\includegraphics[angle=270,scale=0.3]{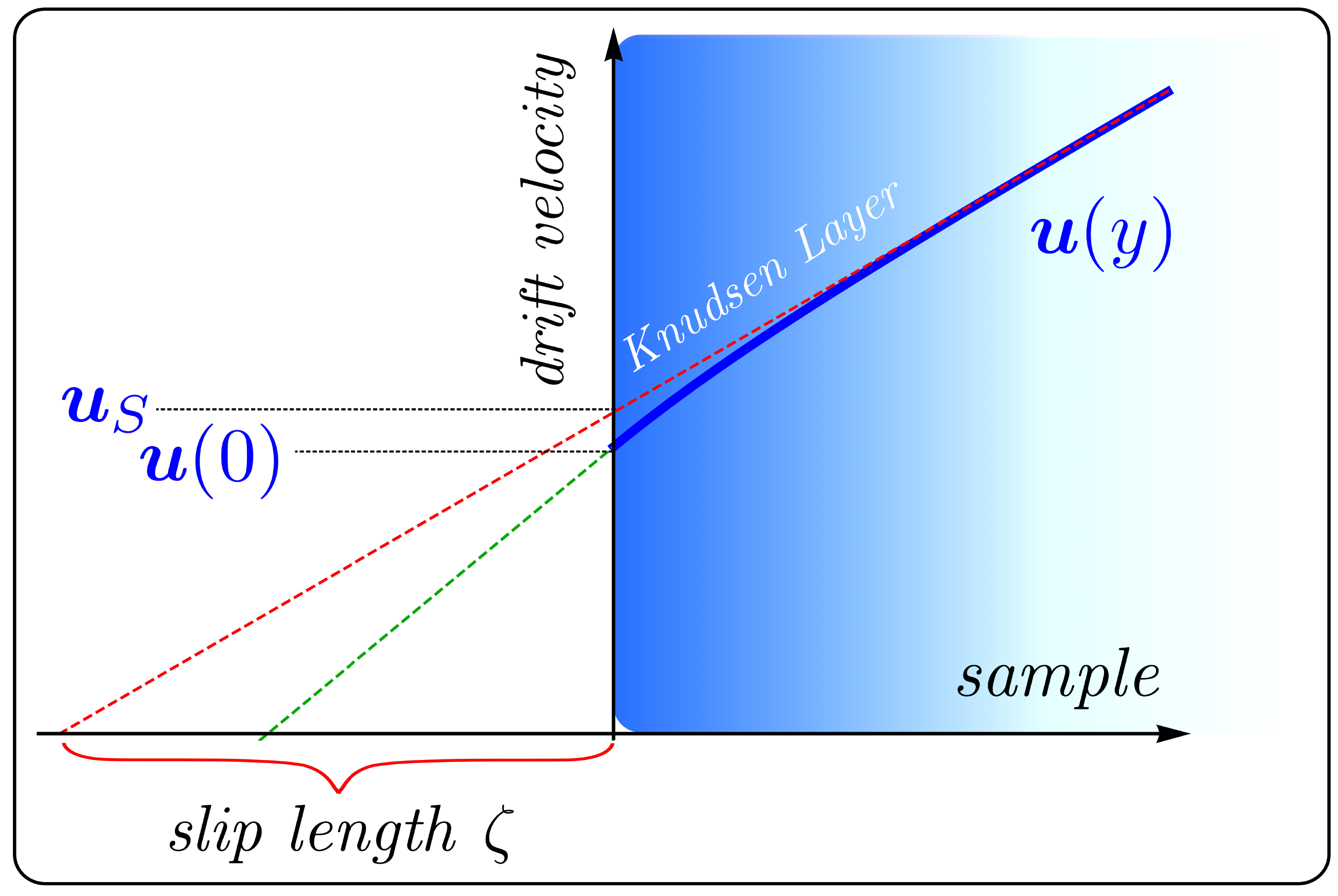}
\par\end{centering}

\caption{The Knudsen-layer - an approximately one mean free path broad region
along the momentum dissipating boundary of a liquid - is depicted.
The fluid behavior in the Knudsen layer is not described by hydrodynamic
equations. The reasoning behind the introduction of the slip length
$\zeta$ is to pass the information about Knudsen layer physics to
the hydrodynamic equations. While the drift velocity $\bm{u}\left(y\right)$
in the Knudsen layer is smaller than the surface velocity $\bm{u}_{S}$
of Eq. (\ref{eq:General_boundary_condition}), the gradient $\partial u_{x}\left(y\right)/\partial y$
approaches the value prescribed by $u_{S}=\partial u/\partial y|_{S}$
at the very end of the layer. Extrapolating this gradient up to zero
velocity one obtains the slip length $\zeta$.\label{fig:Knudsen-layer}}
\end{figure}
In addition, one assumes that $f\left(0\right)=f^{0}$ for positive
velocities $v_{y}>0$, which is equivalent to
\begin{equation}
g\left(0\right)=p_{x}u\left(0\right)\frac{\partial f^{0}}{\partial\epsilon},
\end{equation}
again holding for $v_{y}>0$. With these boundary conditions, (\ref{eq:BGK_kin_eq})
is solved by
\begin{align}
g_{v_{y}>0} & =p_{x}u\left(0\right)\frac{\partial f^{0}}{\partial\epsilon}e^{-\frac{y}{\tau v_{y}}}+\int_{0}^{y}dy'p_{x}u'\left(y'\right)\frac{\partial f^{0}}{\partial\epsilon}e^{-\frac{\left|y'-y\right|}{\tau v_{y}}}\nonumber \\
g_{v_{y}<0} & =-\int_{y}^{\infty}dy'p_{x}u'\left(y'\right)\frac{\partial f^{0}}{\partial\epsilon}e^{\frac{\left|y'-y\right|}{\tau v_{y}}}.
\end{align}
The influx current of tangential momentum into the Knudsen layer is
given by $-\eta u'$$\left(\infty\right)$. The authors of Ref \cite{HJensen1980}
further assume that this current is constant in the whole Knudsen
layer, and only at the boundary is it converted into a tangential
flow that creates a velocity slip. In our treatment of the kinetic
distribution at the nearly specular boundary in section \ref{sub:Kinetic-equation-at-boundary},
where we showed that the variation of the tangential momentum current
gives a contribution subleading in the small parameter $h^{2}L/\lambda_{T}^{3}$,
we explicitly saw that this assumption holds. If it holds in the totally
diffuse case as well, we can write 
\begin{align}
-\eta u'\left(\infty\right) & =\int_{>}\frac{d^{d}k}{\left(2\pi\right)^{d}}v_{y}p_{x}^{2}u\left(0\right)\frac{\partial f^{0}}{\partial\epsilon}e^{-\frac{y}{\tau v_{y}}}\nonumber \\
 & +\int_{>}\frac{d^{d}k}{\left(2\pi\right)^{d}}v_{y}p_{x}^{2}\int_{0}^{y}dy'u'\left(y'\right)\frac{\partial f^{0}}{\partial\epsilon}e^{-\frac{\left|y'-y\right|}{\tau v_{y}}}\nonumber \\
 & -\int_{<}\frac{d^{d}k}{\left(2\pi\right)^{d}}\int_{y}^{\infty}dy'v_{y}p_{x}^{2}u'\left(y'\right)\frac{\partial f^{0}}{\partial\epsilon}e^{\frac{\left|y'-y\right|}{\tau v_{y}}}.\label{eq:BGK_mom_curr_balance}
\end{align}
Eq. (\ref{eq:BGK_mom_curr_balance}) is an integral equation for $u'\left(y\right)$.
Reference \cite{HJensen1980} develops a method to extract from Eq.
(\ref{eq:BGK_mom_curr_balance}) information about the slip length,
without seeking an explicit solution: First, the function $L_{n}\left(y\right)$
is introduced such that
\begin{equation}
L_{n}\left(y\right)=g_{N}\int_{>}\frac{d^{d}k}{\left(2\pi\right)^{d}}v_{y}p_{x}^{2}\left(\tau v_{y}\right)^{n-1}\left(-\frac{\partial f^{0}}{\partial\epsilon}\right)e^{-\frac{y}{\tau v_{y}}}.
\end{equation}
$g_{N}$ accounts for additional degeneracies. In the case of graphene,
a factor of $g_{N}=2N$ in front of the integral will account for
excitations with positive and negative energies and the spin-valley
degeneracy. The viscosity can be expressed as 
\begin{equation}
\eta=\tau\int\frac{d^{d}k}{\left(2\pi\right)^{d}}v_{y}^{2}p_{x}^{2}\left(-\frac{\partial f^{0}}{\partial\epsilon}\right)=2L_{2}\left(0\right).
\end{equation}
Definig the function $\Psi\left(z\right)$ via the equation $u'\left(z\right)=u'\left(\infty\right)$$\left(1+\Psi\left(z\right)\right)$
and introducing $y_{0}=u\left(0\right)/u'\left(\infty\right)$, one
can reduce Eq. (\ref{eq:BGK_mom_curr_balance}) to
\begin{equation}
z_{0}L_{1}\left(z\right)-L_{2}\left(z\right)=-\int_{0}^{\infty}dz'\Psi\left(z'\right)L_{1}\left(\left|z-z'\right|\right).\label{eq:BGK_int_eq}
\end{equation}
Notice that since the drift velocity is expected to drop compared
to the hydrodynamic boundary value $u_{S}=u'\left(\infty\right)\zeta$,
the function $\Psi\left(z\right)$ is expected to be positive everywhere
and to vanish for $y\rightarrow\infty$. For the slip length $\zeta$
the following holds (see Fig. \ref{fig:Knudsen-layer}):
\begin{align}
u_{S}=u'\left(\infty\right)\zeta & =u\left(0\right)+\int_{0}^{\infty}u'\left(z'\right)-u'\left(\infty\right)dy'\nonumber \\
\zeta & =y_{0}+\int_{0}^{\infty}\Psi\left(y'\right)dy'.
\end{align}
Together with (\ref{eq:BGK_int_eq}) this yields
\begin{equation}
\zeta L_{1}\left(y\right)-L_{2}\left(y\right)=-\int_{0}^{\infty}dy'\Psi\left(y'\right)\left(L_{2}\left(\left|y-y'\right|\right)-L_{1}\left(y\right)\right).\label{eq:BGK_lower_eq}
\end{equation}
One property of the functions $L_{n}\left(y\right)$ is $dL_{n}\left(y\right)/dy=-L_{n-1}\left(y\right)$.
Using this relationship, the above equation can be integrated over
the region $y>0$ to yield
\begin{equation}
\zeta L_{2}\left(0\right)-L_{3}\left(0\right)=\int_{0}^{\infty}dy'\Psi\left(y'\right)\left(L_{2}\left(y'\right)-L_{2}\left(0\right)\right).\label{eq:BGK_upper_eq}
\end{equation}
Noticing that $L_{n}\left(y>0\right)<L_{n}\left(0\right)$ and remembering
that $\Psi\left(z\right)$ is positive one obtains from (\ref{eq:BGK_lower_eq})
\begin{equation}
\zeta>\frac{L_{2}\left(0\right)}{L_{1}\left(0\right)}\label{eq:BGK_lower_bound}
\end{equation}
and from (\ref{eq:BGK_upper_eq})
\begin{equation}
\zeta<\frac{L_{3}\left(0\right)}{L_{2}\left(0\right)}.\label{eq:BGK_upper_bound}
\end{equation}
The equations (\ref{eq:BGK_lower_bound}) and (\ref{eq:BGK_upper_bound})
consitute a lower and an upper bound on the slip length, which are
typically not too far apart and therefore give a good estimate for
$\zeta$. The lower bound can even be improved with the help of the
inequality $L_{n+1}L_{n-1}>L_{n}^{2}$: Realizing that $d/dy\left(L_{n}\left(y\right)/L_{n+1}\left(y\right)\right)=\left(L_{n}^{2}-L_{n+1}L_{n-1}\right)/L_{n+1}^{\text{2}}<0$,
we have $L_{n}\left(y\right)/L_{n}\left(0\right)>L_{n-1}\left(y\right)/L_{n-1}\left(0\right)$.
The combination of equations (\ref{eq:BGK_lower_eq}) and (\ref{eq:BGK_upper_eq})
then yields
\begin{equation}
2\zeta-\frac{L_{2}\left(0\right)}{L_{1}\left(0\right)}-\frac{L_{3}\left(0\right)}{L_{2}\left(0\right)}=\int_{0}^{\infty}dy'\Psi\left(y'\right)\left(\frac{L_{2}\left(y'\right)}{L_{2}\left(0\right)}-\frac{L_{1}\left(y'\right)}{L_{1}\left(0\right)}\right)
\end{equation}
and
\begin{equation}
\zeta>\frac{1}{2}\left(\frac{L_{2}\left(0\right)}{L_{1}\left(0\right)}+\frac{L_{3}\left(0\right)}{L_{2}\left(0\right)}\right).\label{eq:BGK_improved_lower_bound}
\end{equation}
As realized by the authors of \cite{HJensen1980}, the lower bound
(\ref{eq:BGK_lower_bound}) is equivalent to the slip length of the
Maxwell approach. Remembering the general form of the distribution
function (\ref{eq:General_Chapman}) one easily sees that in the case
of diffuse scattering, and assuming that particles at the boundary
are described by the bulk distribution, Eq. (\ref{eq:Mom_curr_average})
reads
\begin{equation}
0=-\frac{1}{2}\eta\frac{\partial u_{x}}{\partial y}+\int\frac{d^{d}k}{\left(2\pi\right)}v_{y}p_{x}^{2}u\left(0\right)\left(-\frac{\partial f^{0}}{\partial\epsilon}\right),
\end{equation}
where we have ommited the summation over the two graphene bands. Approximating
$\zeta_{lower}=u\left(0\right)/\left(\partial u_{x}/\partial y\right)$
(for the exact slip length we need to replace $u\left(0\right)$ by
$u_{S}$), one obtains
\begin{equation}
\zeta_{lower}=\frac{\eta/2}{\int\frac{d^{d}k}{\left(2\pi\right)}v_{y}p_{x}^{2}\left(-\frac{\partial f^{0}}{\partial\epsilon}\right)}=\frac{L_{2}\left(0\right)}{L_{1}\left(0\right)},\label{eq:BGK_Maxwell_Bound}
\end{equation}
which is the lower bound (\ref{eq:BGK_lower_bound}) and is, of course,
identical to our result of Eq. (\ref{eq:Diffuse_slip_length}). For
graphene, the upper bound (\ref{eq:BGK_upper_bound}) and the lower
bound are very close: $\zeta_{upper}\approx1.15\zeta_{lower}$. A
comparison of the slip length for the classical kinetic gas given
by (\ref{eq:BGK_improved_lower_bound}) and (\ref{eq:BGK_upper_bound})
with exact and numerical results was performed in \cite{HJensen1980}.
The authors report a deviation of less than 1\%. For completeness,
we note that the $\zeta$ obtained from Eq. (\ref{eq:Mom_average})
is equivalent to the lower bound set by $\zeta>L_{1}\left(0\right)/L_{0}\left(0\right)$,
which is worse than the lower bound (\ref{eq:BGK_lower_bound}).

\section{Flow through a strip with a circular obstacle\label{sec:Flow_Strip_Obstacle}}

If the slip length of an electron liquid $\zeta$ is much larger than
the typical sample size, it is appropriate to use the no-stress boundary
condition of Eq. (\ref{eq:no_stress_condition}) to model the interaction
of the liquid with the wall. If this condition is applied, the conductivity
of a clean sample with a Poiseuille geometry is infinite, as is clear
from Eq. (\ref{eq:Poiseuille_current}). However, if viscous shear
forces act somewhere in the sample, the conductivity becomes finite.
This can be used to identify hydrodynamic flow, even when the Gurzhi
effect should not be observable at large $\zeta$. Viscous shear forces
arise, for example, if the fluid has to bypass an impenetrable obstacle
that is put somewhere in the sample. As an illustration, we consider
a graphene strip that is infinitely extended in the $x$-direction
and goes from $y=-w/2$ to $y=w/2$. The obstacle shall be a disk
of radius $a$ placed at the origin of the coordinate system. No-stress
boundary conditions shall apply at the interface of obstacle and liquid
as well as at $y=\pm w/2$. We calculate the pressure difference that
arises due to viscous shear forces between the ends of the strip at
$x=-\infty$ and $x=\infty$. In what follows, graphene at charge
neutrality is considered but the calculation can be readily modified
to suit the Fermi liquid case. In the former case the flow should
be probed using thermal transport, while it is given by the electrical
current in the latter case.

\begin{figure}
\centering{}\includegraphics[scale=0.6]{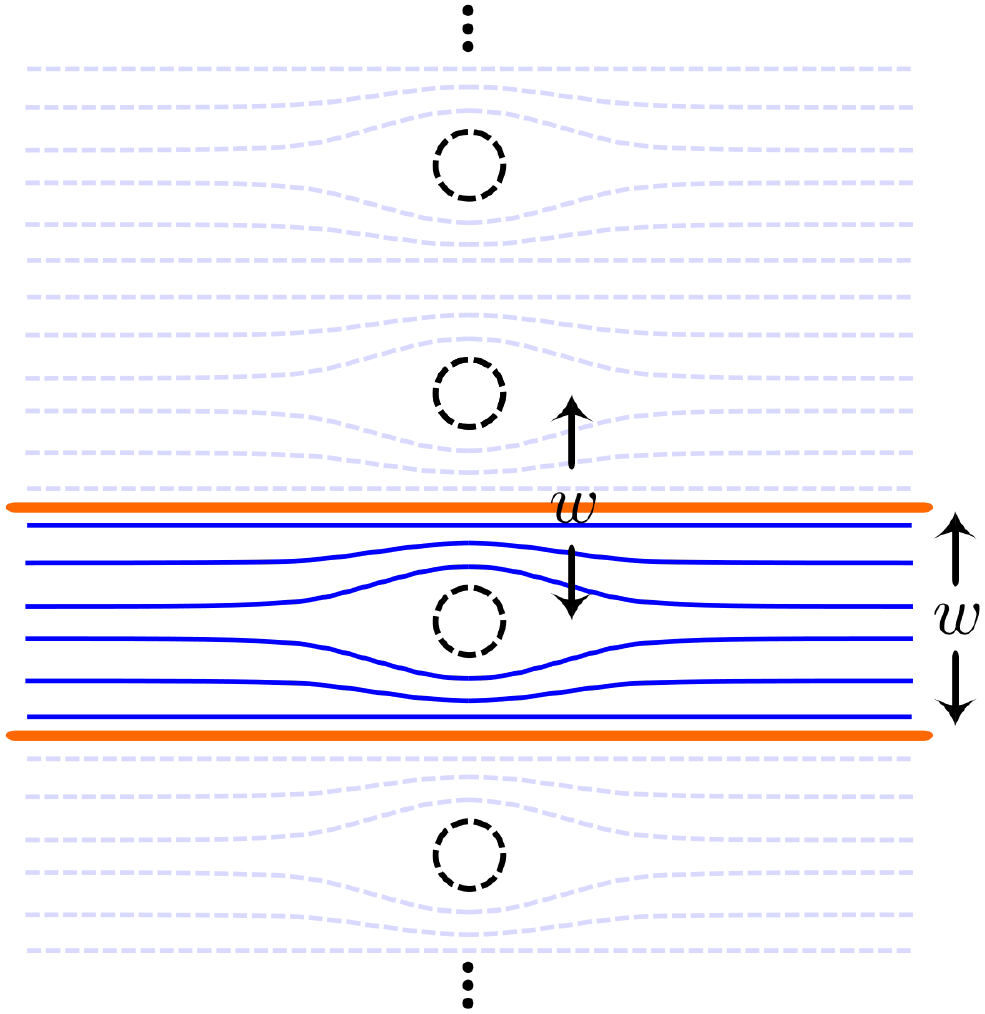}\caption{The method of images can be used to solve Eq. (\ref{eq:Oseen's_equation})
for a strip-like sample geometry with a circular obstacle. The sample
and the electron flow are drawn in color, the image fields are shown
in light grey. Infinitely many image solutions must be placed symmetrically
around the obstacle in the middle of the sample. \label{fig:Images}}
\end{figure}

\begin{figure}
\centering{}\includegraphics[angle=270,scale=0.4]{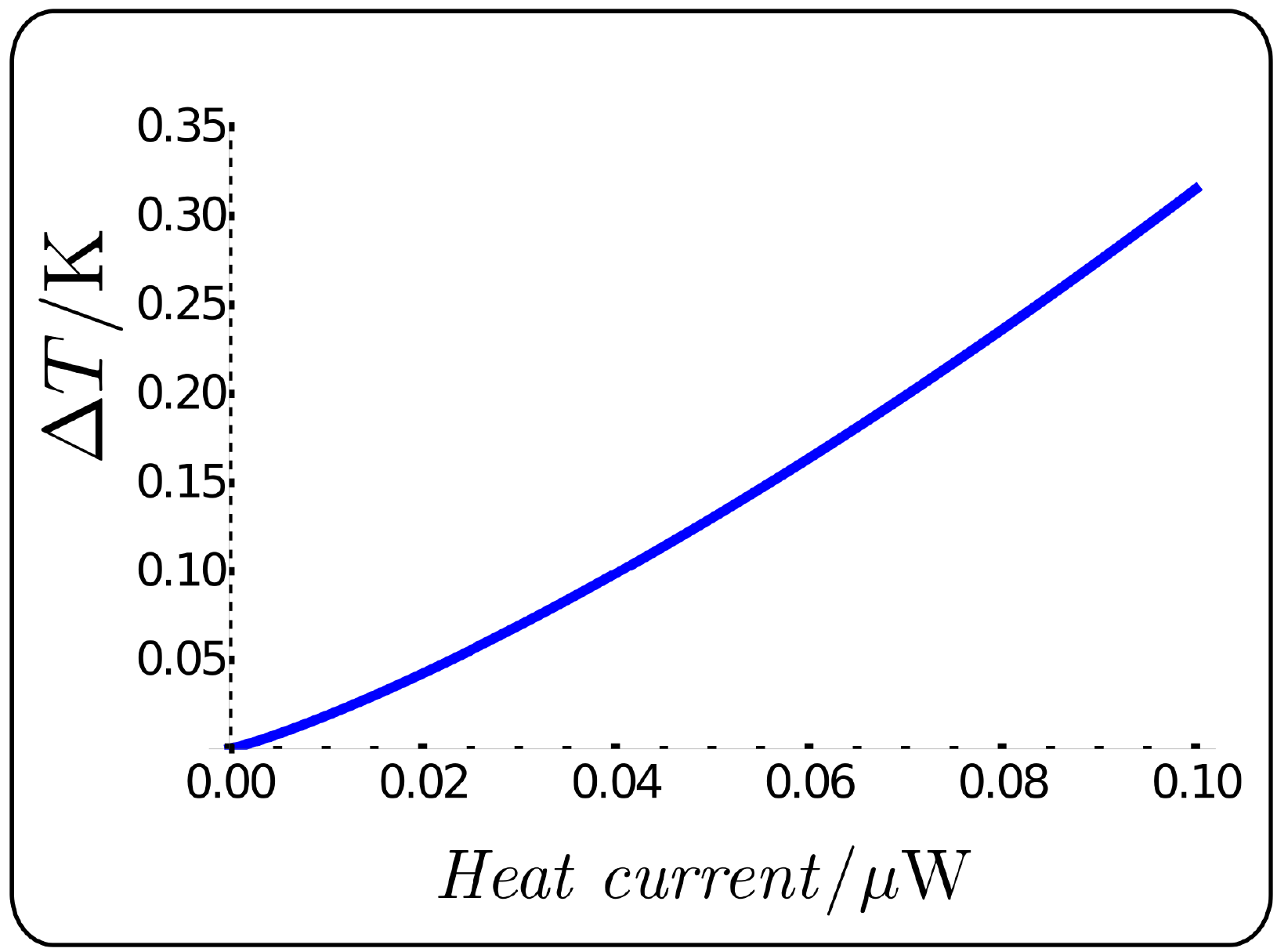}\caption{Temperature difference induced by a heat current $I_{E}$ through
a $10\,\upmu\textrm{m}$ wide strip of charge neutral graphene at
$50\,\textrm{K}$ with a circular obstace of radius $1\,\upmu\textrm{m}$
at the center of the strip (see Fig. \ref{fig:Images}). At a heat
current of $0.1\upmu\textrm{m}$ the Reynolds number is $\sim0.74$,
approaching unity.\label{fig:temperature_heat_current}}
\end{figure}

\begin{figure}
\centering{}\includegraphics[angle=270,scale=0.4]{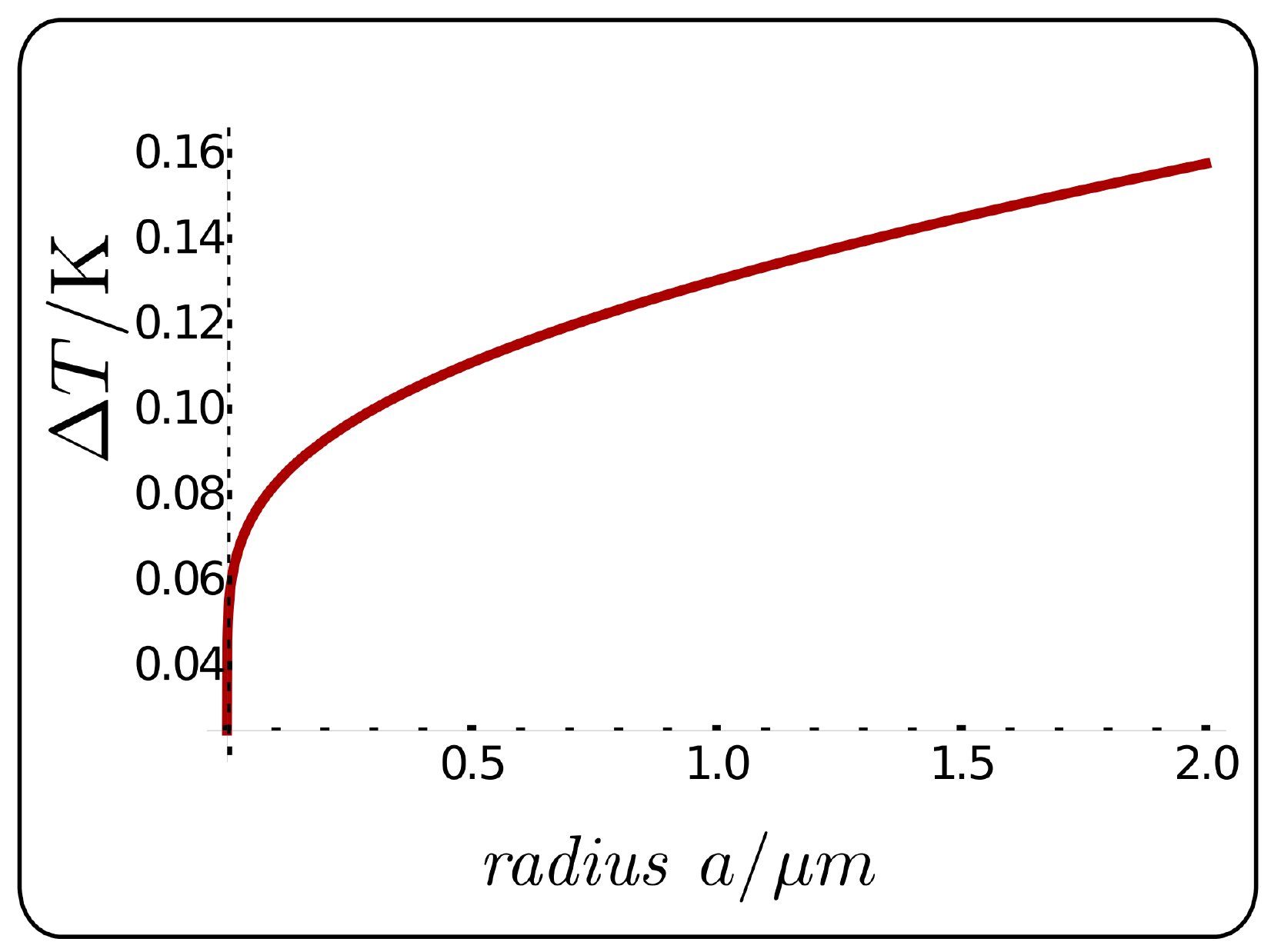}\caption{Temperature difference for a fixed heat current of $I_{E}=0.5\upmu\textrm{W}$
through a $10\,\upmu\textrm{m}$ wide graphene channel at $50K$ as
a function of the obstacle radius $a$. \label{fig:temperature_radius}}
\end{figure}

The full Navier-Stokes equation for graphene electrons reads \cite{Mueller2009,Briskot2015}
\begin{equation}
\frac{\tilde{w}}{v^{2}}\left(\partial_{t}\bm{u}+\left(\bm{u}\cdot\nabla\right)\bm{u}\right)+\nabla p+\frac{\partial_{t}p}{v^{2}}\bm{u}-\eta\nabla^{2}\bm{u}=0.\label{eq:Full_NS-eq}
\end{equation}
$\tilde{w}$ is the enthalpy density $\tilde{w}=\frac{3n_{E}/2}{2+\left|\bm{u}\right|/v}\approx\tilde{w}_{0}=\frac{3}{2}n_{E}$.
To begin with, we consider a liquid which is not bounded at $y=w/2$.
As is well known \cite{Batchelor,Lucas2017}, in two dimensions the
flow around a circular obstacle exhibits Stokes' paradox: the flux
$\bm{u}$ is not a linear function of $\nabla p$ for small $\nabla p$.
The usual way to circumvent this problem is to use Oseen's equation
\cite{Batchelor} in which the flow $\bm{u}$ is linearized around
a spatially constant flow $\bm{U}\propto\hat{\bm{e}}_{x}$. The full
flow $\bm{u}$ is then written
\begin{equation}
\bm{u}=\bm{U}+\bm{q},
\end{equation}
and the linearized Navier-Stokes equation reads
\begin{equation}
\frac{\tilde{w}_{0}}{v^{2}}\left(\bm{U}\cdot\nabla\right)\bm{q}+\nabla p-\eta\nabla^{2}\bm{q}=0.\label{eq:Oseen's_equation}
\end{equation}
In Appendix \ref{sec:Flow_Infinite_Domain} we give the general solution
to Eq. (\ref{eq:Oseen's_equation}) following the analysis of Ref.
\cite{Tomotika1950}. We also calculate the flow around the obstacle
for an arbitrary $\zeta$ on an infinite domain. If the flow is not
confined to the strip, the pressure induced by the obstacle vanishes
at infinity, where $p\sim1/r$. If, however, the flow is bounded at
$y=\pm w/2$, the obstacle does induce a pressure difference along
the strip. The boundary conditions imposed on the electron flow by
the two walls at $y=\pm w/2$ are 
\begin{eqnarray}
q_{y}\left(y=\pm w/2\right) & = & 0\nonumber \\
\frac{\partial q_{x}\left(y=\pm w/2\right)}{\partial y} & = & 0.\label{eq:strip_conditions}
\end{eqnarray}
These boundary conditions can be implemented using the method of images,
known from electrostatics. The expression 
\begin{align}
\bm{q}_{tot} & =\sum_{j=-\infty}^{\infty}\bm{q}\left(x,y+jw\right),
\end{align}
with $\bm{q}\left(x,y\right)$ being the infinite domain solution
obtained in Eq. (\ref{eq:q_gen_sol})-(\ref{eq:B0_eq}), does satisfy
Eq. (\ref{eq:Oseen's_equation}) everywhere inside the strip and matches
the conditions of Eq. (\ref{eq:strip_conditions}). It corresponds
to infinitely many image fields placed along the $y$-axis, symmetrically
to the original obstacle at $y=0$ (see Fig. \ref{fig:Images}). The
solution $\bm{q}_{tot}$ is only approximate, since the boundary condition
at the surface of the obstacle is not matched exactly. It is matched,
however, at $r=a$ and $y=0$. Therefore, the error is of order $\sim\frac{a}{w}$
, i. e. small, if the obstacle is small compared to the width of the
strip.
The pressure distribution along the sample can then be calculated
from the function
\begin{equation}
\phi_{j}=\phi\left(x,y+jw\right),
\end{equation}
which solves the Laplace equation and is defined in Eq. (\ref{eq:pressure_phi}).
The pressure generated by every single image field is $p\left(x,y+jm\right)=U\left(\tilde{w}_{0}/v^{2}\right)\partial\phi_{j}/\partial x$
(for details see Appednix \ref{sec:Flow_Infinite_Domain} and Ref
\cite{Tomotika1950}). The total pressure at $y=0$ is 
\begin{eqnarray}
p_{tot} & = & \sum_{j=-\infty}^{j=\infty}p\left(x,jm\right)\nonumber \\
 & = & \frac{\pi A_{0}U^{2}}{w}\frac{\tilde{w_{0}}}{v^{2}}\coth\left(\frac{\pi x}{w}\right)\nonumber \\
 &  & +\text{sgn}(x)\frac{\pi^{2}A_{1}U^{2}}{w^{2}}\frac{\tilde{w_{0}}}{v^{2}}\textrm{sinh}\left(\frac{\pi x}{w}\right)^{-2}.
\end{eqnarray}
The constants $A_{0}$ and $A_{1}$ are given in Eqs. (\ref{eq:A0_B0})
and (\ref{eq:A1_B0}). While the pressure of any single image field
$p\left(x,jm\right)$ vanishes for $x\rightarrow\pm\infty$, the sum
over all image fields remains finite. The pressure difference across
the sample is then
\begin{equation}
\Delta p=p_{tot}\left(x\rightarrow\infty\right)-p_{tot}\left(x\rightarrow-\infty\right)=2\frac{\pi A_{0}U^{2}}{w}\frac{\tilde{w_{0}}}{v^{2}}.
\end{equation}
Using Eqs. (\ref{eq:A0_B0}), (\ref{eq:B0_eq}) and expanding $B_{0}$
for small $U$ (small Reynolds numbers), as well as taking the limit
$\zeta\rightarrow\infty$ for the slip length at the obstacle, we
obtain
\begin{equation}
\Delta p=-\frac{8\pi U}{3-2\left(\log\left(\frac{aU}{4\nu}\right)+\gamma\right)}\frac{\tilde{w_{0}}}{v^{2}}\frac{\nu}{w}.\label{eq:Pressure_and_Slow_Flow}
\end{equation}
$\nu$ is the kinematic viscosity $\nu=\left(v^{2}/\tilde{w}_{0}\right)\eta$.
As expected, the pressure arising due to a small flow velocity $U$
cannot be linearized, which is a manifestation of Stoke's paradox.
We want to link this result to an experimental setup in which a heat
flow through the sample will induce a temperature difference. With
the help of Eq. (\ref{eq:pressure_temperature_relation}) we can rewrite
the pressure difference as a temperature difference. The flow velocity
$U$ is connected to the heat current density through the formula
\cite{Briskot2015} 
\begin{equation}
\bm{j}_{E}=\frac{3n_{E}\bm{U}}{2+U^{2}/v^{2}}\approx\frac{3}{2}n_{E}\bm{U}.
\end{equation}
With the total energy current being $I_{E}=j_{E}w$ we can write for
Eq. (\ref{eq:Pressure_and_Slow_Flow})
\begin{equation}
\left|\Delta T\right|=\frac{16\pi I_{E}\eta/\left(n_{E}w^{2}s\right)}{9-6\left(\log\left(\frac{1}{9}\frac{I_{E}a}{v^{2}\eta w}\right)+\gamma\right)}.\label{eq:induced_temperature_difference_dynvisc}
\end{equation}
$\gamma\approx0.58$ is the Euler constant. This result can be used
to determine the viscosity $\eta$. The entropy density is given by
\cite{Mueller2009}
\[
s=N\frac{9\zeta\left(3\right)}{\pi}k_{B}\frac{\left(k_{B}T\right)^{2}}{\hbar^{2}v^{2}}.
\]
Fig. \ref{fig:temperature_heat_current} shows the dependece of the
induced temperature difference on the heat current through a $10\,\upmu\textrm{m}$
wide graphene sample at $50\,\textrm{K}$ for an obstacle of radius
$a=1\,\upmu\textrm{m}$. The dependence of the temperature difference
$\left|\Delta T\right|$ on the radius $a$ is shown in Fig. \ref{fig:temperature_radius}.
The scaling behavior of the current with $a$ and $w$ is non-trivial
due to the presence of a third length scale $\nu/U$. Since no momentum
is dissipated at the sample boundaries the temperature difference
along the sample, far enough from the disc, does not depend on the
length of the sample. The temperature difference is induced in the
region near the disc only.

\section{Conclusions}

Hydrodynamic flow sensitively depends on the nature of the boundary
conditions for the velocity flow field. These boundary conditions
can efficiently be characterized by the slip length $\zeta$ introduced
in Eq. (\ref{eq:General_boundary_condition}). In order to obtain
a quantitative understanding of the slip length in electron fluids,
we have derived the slip lengths at different kinds of edges for Dirac
and Fermi liquids. We found that for viscous electronic flow the slip
length can always be written in the form 
\begin{equation}
\zeta=f\left(\kappa\right)l_{ee}
\end{equation}
with dimensionless ratio $\kappa=h^{2}h'^{d-1}/\lambda^{d+1}$ . $\kappa$
depends on the two length scales $h$ and $h'$ that characterize
the interface scattering and the electron wavelength $\lambda$, respectively.
For graphene at the neutrality point, the latter is strongly temperature
dependent ($\lambda=\hbar v/(k_{{\rm B}}T)$), while it corresponds
to the Fermi wave length in the case of Fermi liquids ($\lambda=1/k_{{\rm F}}$).
The dimensionless function $f\left(\kappa\right)$ diverges for small
$\kappa$: $f\left(\kappa\ll1\right)=f_{0}/\kappa$, while it approaches
a constant for strong interface scattering: $f\left(\kappa\rightarrow\infty\right)\rightarrow f_{\infty}$.
We determined $f_{\infty}$ using the assumption of diffuse scattering.
The numerical values for the coefficients $f_{0}$ and $f_{\infty}$
depend sensitively on the electronic dispersion relation and dimensionality
of the system.

Since for all quantum fluids the mean free path diverges as the temperature
approaches zero, the ultimate behavior of the slip length at low temperatures
is $\zeta\rightarrow\infty$ and the no-stress boundary conditions
are appropriate. For Dirac fluids in samples with weakly disordered
edges even the ratio $\zeta/l_{ee}$ diverges as $T\rightarrow0$.
At intermediate temperatures the slip lengths are such that no-slip
boundary conditions may be justified for large sample sizes. In particular,
we show that the electron viscosity of PdCoO$_{2}$ \cite{Moll2016}
and WP$_{2}$ \cite{Gooth2017} is small enough, such that Poiseuille
type flow can manifest itself, as seen experimentally. The linear
Dirac spectrum and the typical sample sizes used imply that graphene
is essentially always in the regime of no-stress conditions. If no-stress
boundary conditions apply, it is no longer possible to detect Poiseuille
type flow and the Gurzhi effect. However, hydrodynamic effects such
as superballistic flow \cite{KrishnaKumar2017} and the negative local
resistivity due to vorticity can still be observed \cite{Bandurin}.
In addition we propose the flow through a channel with a circular
obstacle as an efficient approach to identify hydrodyanmic flow. Thus,
one of the most characteristic features of the hydrodynamics of electron
fluids is the nature of the boundary condition of the flow velocity.
The fact that for a broad range of parameters the slip lengths of
quantum fluids are very large makes electron hydrodynamics distinct
from its well studied classical counterpart.

\section{Acknowledgement}

We thank I. V. Gornyi, V. Yu. Kachorovskii, A. D. Mirlin, B. N. Narozhny,
D. G. Polyakov and P. W\"olfle for stimulating discussions.

\appendix
\numberwithin{equation}{section}

\section{Pressure and temperature gradients in charge neutral graphene}

In this appendix, we show how pressure gradients can be related to
temperature gradients in graphene at charge neutrality. From the Gibbs-Duhem
relation we know that the pressure of a system is equal to minus the
grand potential density
\[
\frac{\Omega}{V}=-p.
\]
The standard expression for $\Omega/V$ can be integrated by parts
to give
\begin{eqnarray*}
\frac{\Omega}{V} & = & -\beta^{-1}\sum_{\lambda}\int\frac{d^{2}k}{\left(2\pi\right)^{2}}\ln\left(1+e^{-\beta\left(\epsilon_{\lambda,\bm{k}}-\mu\right)}\right)\\
 & = & \beta^{-1}\sum_{\lambda}\int\frac{d^{2}k}{\left(2\pi\right)^{2}}k_{i}\frac{\partial}{\partial k_{i}}\ln\left(1+e^{-\beta\left(\epsilon_{\lambda,\bm{k}}-\mu\right)}\right)\\
 &  & \quad-\frac{1}{4\pi}\Lambda^{2}\left(v\Lambda-\mu\right).
\end{eqnarray*}
An upper cut-off $\Lambda$ for the momentum integration over the
$\lambda=-1$ band was introduced. We therefore have
\begin{eqnarray*}
p & = & \sum_{\lambda}\int\frac{d^{2}k}{\left(2\pi\right)^{2}}\frac{\lambda v_{i}k_{i}}{1+e^{\beta\left(\epsilon_{\lambda,\bm{k}}-\mu\right)}}\\
 &  & \quad+\frac{1}{4\pi}\Lambda^{2}\left(v\Lambda-\mu\right).
\end{eqnarray*}
The first right hand side term is the expression for pressure $p_{kin}$
as it enters the kinetic theory, the second term is the Fermi pressure
$p_{\Lambda}$ of the occupied lower Dirac cone:
\[
p=p_{kin}+p_{\Lambda}.
\]
The pressure gradient can be written
\[
\nabla p_{kin}=\frac{\partial p_{kin}}{\partial T}\nabla T+\frac{\partial p_{int}}{\partial\mu}\nabla\mu.
\]
Since $p_{\Lambda}$ does not depend on temperature, the relation
\[
s=-\frac{\partial\left(\Omega/V\right)}{\partial T}=\frac{\partial p_{kin}}{\partial T}
\]
holds, where $s$ is the entropy density. On the other hand, at charge
neutrality ($\mu=0$) we have
\begin{eqnarray*}
\left.\frac{\partial p_{kin}}{\partial\mu}\right|_{\mu=0} & = & -\sum_{\lambda}\int\frac{d^{2}k}{\left(2\pi\right)^{2}}\left(\lambda v_{i}k_{i}\right)\frac{\partial}{\partial\epsilon_{\lambda,\bm{k}}}\frac{1}{e^{\beta\epsilon_{\lambda,\bm{k}}}+1}\\
 & = & \sum_{\lambda}\int\frac{d^{2}k}{\left(2\pi\right)^{2}}\frac{1}{e^{\beta\epsilon_{\lambda,\bm{k}}}+1}-\frac{1}{4\pi}\Lambda^{2},
\end{eqnarray*}
where again the last term stems from the integration boundary at $\lambda=-1$,
$k\rightarrow\infty$. Being aware that
\begin{eqnarray*}
\sum_{\lambda}\int\frac{d^{2}k}{\left(2\pi\right)^{2}}\frac{1}{e^{\beta\epsilon_{\lambda,\bm{k}}}+1} & = & \int_{0}^{2\pi}\frac{d\varphi}{\left(2\pi\right)^{2}}\int^{\Lambda}kdk\,1,
\end{eqnarray*}
we have
\[
\left.\frac{\partial p_{kin}}{\partial\mu}\right|_{\mu=0}=0,
\]
and therefore
\[
\nabla p_{kin}=s\nabla T.
\]
For simplicity, in the main text we refer to $p_{kin}$ as $p$.

\section{Bulk distribution function for graphene at charge neutrality}

In what follows we summarize the main steps of the calculation of
the shear viscosity of graphene originally determined in Ref. \cite{Mueller2009}.
In addition to the analysis presented in Ref. \cite{Mueller2009},
we also show the behavior at finite frequency. The Hamiltonian for
electrons in graphene that interact via the long-range Coulomb repulsion
consists of the noninteracting part 
\begin{equation}
H_{0}=v\hbar\int_{k}\sum_{\alpha\beta,i}\psi_{ai}^{\dagger}\left(\mathbf{k}\right)\left(\mathbf{k\cdot\sigma}\right)_{ab}\psi_{bi}\left(\mathbf{k}\right)
\end{equation}
and the interaction 
\begin{equation}
H_{\mathrm{int}}=\frac{1}{2}\int_{k,k^{\prime},q}\sum_{ab,ij}V\left(\mathbf{q}\right)\psi_{\mathbf{k+q,}a,i}^{\dagger},\psi_{\mathbf{k}^{\prime}-\mathbf{q,}b,j}^{\dagger}\psi_{\mathbf{k}^{\prime},b,j}\psi_{\mathbf{k,}a,i}
\end{equation}
with $V\left(q\right)=\frac{2\pi e^{2}}{\varepsilon\left\vert \mathbf{q}\right\vert }$.
Here $i=1,...,N=4$ refers to the spin and valley flavors. $H_{0}$
is diagonalized by a unitary transformation $U_{\mathbf{k}}$. The
eigenvalues of $H_{0}$ are $\varepsilon_{\mathbf{k\lambda}}=\pm v\hbar k$
where $k=\left\vert \mathbf{k}\right\vert $. The quasiparticle states
for the two bands are $\gamma_{\mathbf{k}}=U_{\mathbf{k}}\psi_{\mathbf{k}}$,
with

\begin{equation}
H_{0}=v\hbar\int_{\mathbf{k}}\sum_{\lambda=\pm,i}\lambda k\gamma_{\mathbf{k},\lambda,i}^{\dagger}\gamma_{\mathbf{k},\lambda,i}.
\end{equation}
In the band representation follows for the Coulomb interaction 
\begin{eqnarray}
H_{\mathrm{int}} & = & \frac{1}{2}\int_{k,k^{\prime},q}\sum_{\lambda\mu\mu^{\prime}\lambda^{\prime},ij}T_{\lambda\mu\mu^{\prime}\lambda^{\prime}}\left(\mathbf{k,k}^{\prime},\mathbf{q}\right)\nonumber \\
 &  & \times\gamma_{\mathbf{k+q,}\lambda^{\prime},i}^{\dagger}\gamma_{\mathbf{k}^{\prime}-\mathbf{q,}\mu,j}^{\dagger}\gamma_{\mathbf{k}^{\prime},\mu^{\prime},j}\gamma_{\mathbf{k,}\lambda,i}
\end{eqnarray}
where 
\begin{equation}
T_{\lambda\mu\mu^{\prime}\lambda^{\prime}}\left(\mathbf{k,k}^{\prime},\mathbf{q}\right)=V\left(q\right)\left(U_{\mathbf{k+q}}U_{\mathbf{k}}^{-1}\right)_{\lambda^{\prime}\lambda}\left(U_{\mathbf{k}^{\prime}-\mathbf{q}}U_{\mathbf{k}^{\prime}}^{-1}\right)_{\mu\mu^{\prime}}.
\end{equation}
The goal is to determine the distribution function

\begin{equation}
f_{\mathbf{k}\lambda}\left(\mathbf{x},t\right)=\left\langle \gamma_{\mathbf{k},\lambda}^{\dagger}\gamma_{\mathbf{k},\lambda}\right\rangle _{\mathbf{x},t}
\end{equation}
for a state with momentum $\mathbf{k}$ and band index $\lambda$
at position $\mathbf{x}$. To this end we solve the Boltzmann equation
\begin{equation}
\frac{\partial f_{\mathbf{k}\lambda}\left(\mathbf{x},t\right)}{\partial t}+\mathbf{v}_{\mathbf{k},\lambda}\cdot\frac{\partial f_{\mathbf{k}\lambda}\left(\mathbf{x},t\right)}{\partial\mathbf{x}}=-C_{{\rm e.e.\mathbf{k}}\lambda}\left(\mathbf{x},t\right)\label{Boltz}
\end{equation}
in the bulk of the sample. $\mathbf{v}_{\mathbf{k},\lambda}=\frac{1}{\hbar}\frac{\partial\varepsilon_{\mathbf{k\lambda}}}{\partial\mathbf{k}}=\lambda v\frac{\mathbf{k}}{k}$
is the single particle velocity. The collision integral is 
\begin{equation}
C_{{\rm e.e.}\mathbf{k}\lambda}\left(\mathbf{x},t\right)=i\Sigma_{\mathbf{k}\lambda}^{<}\left(\varepsilon_{\lambda}\left(\mathbf{k}\right)\right)\left(1-f_{\mathbf{k}\lambda}\right)+i\Sigma_{\mathbf{k}\lambda}^{>}\left(\varepsilon_{\lambda}\left(\mathbf{k}\right)\right)n_{\mathbf{k}\lambda}.
\end{equation}
where $\Sigma_{\mathbf{k}\lambda}^{\gtrless}\left(\omega\right)$
are the diagonal elements of the self energies for occupied and unoccupied
states, respectively \cite{Kadanoff1962}.

The distribution function is then determined from the kinetic equation
using the Chapman-Enskog approach \cite{Kashuba2008,Fritz2008,Mueller2009}.
If the system flows with velocity $\mathbf{u}\left(\mathbf{x}\right)$,
it holds in the laboratory frame 
\begin{equation}
f_{\mathbf{k}\lambda}\left(\mathbf{x},t\right)=\frac{1}{e^{\beta\left(\varepsilon_{\mathbf{k\lambda}}-\hbar\mathbf{k\cdot u}\left(\mathbf{x}\right)\right)}+1}+\delta f_{\mathbf{k}\lambda}\left(\mathbf{x},t\right).
\end{equation}
The driving force for the shear viscosity is the velocity gradient:
\begin{equation}
X_{\alpha\beta}=\frac{\partial u_{\alpha}}{\partial x_{\beta}}+\frac{\partial u_{\beta}}{\partial x_{\alpha}}-\delta_{\alpha\beta}\nabla\cdot\mathbf{u.}\label{drive}
\end{equation}
To leading order in the velocity gradients follows 
\begin{equation}
\frac{\partial f_{\mathbf{k}\lambda}\left(\mathbf{x},t\right)}{\partial t}+\frac{\lambda\beta\hbar vk}{2^{3/2}}\frac{e^{\beta\hbar vk}I_{\alpha\beta}\left(\mathbf{k}\right)X_{\alpha\beta}}{\left(e^{\beta\hbar vk}+1\right)^{2}}=-C_{\lambda}\left(\mathbf{k},t\right)\label{drift2}
\end{equation}
with $I_{\alpha\beta}\left(\mathbf{k}\right)=\sqrt{2}\left(\frac{k_{\alpha}k_{\beta}}{k^{2}}-\frac{1}{2}\delta_{\alpha\beta}\right)$.
To solve the linearized Boltzmann equation we make the ansatz in the
rest frame of the fluid:

\begin{equation}
f_{\mathbf{k}\lambda}\left(\mathbf{x},t\right)=f^{0}\left(\lambda vk\right)+\frac{\lambda\beta\hbar}{2^{3/2}}\frac{e^{\beta\hbar vk}I_{\alpha\beta}\left(\mathbf{k}\right)X_{\alpha\beta}}{\left(e^{\beta\hbar vk}+1\right)^{2}}\psi\left(\beta\hbar vk,\beta\hbar\omega\right)\label{lin1}
\end{equation}
with Fermi function $f^{0}\left(\epsilon\right)=\frac{1}{e^{\beta\epsilon}+1}$.
Similarly, $\psi\left(K,\Omega\right)$ is a dimensionless function
with dimensionless arguments $K=\frac{\hbar vk}{k_{B}T}$ and $\Omega=\frac{\hbar\omega}{k_{B}T}$.
As shown in Ref. \cite{Mueller2009}, $\psi\left(K,\Omega\right)$
is determined by the linearized Boltzmann equation 
\begin{equation}
\frac{e^{K}}{\left(e^{K}+1\right)^{2}}\left(-i\Omega\psi_{\alpha\beta}\left(K,\Omega\right)+KI_{\alpha\beta}\left(\mathbf{K}\right)\right)=2\pi\alpha^{2}J_{\alpha\beta}\left(\mathbf{K}\right)\label{eq:Linearized_Boltzmann}
\end{equation}
where$\psi_{\alpha\beta}\left(K,\Omega\right)=\psi\left(K,\Omega\right)I_{\alpha\beta}\left(\mathbf{K}\right)$
and is not to be confused with the creation and annihilation operators
$\psi_{ai}^{\dagger}\left(\mathbf{k}\right)$, $\psi_{ai}\left(\mathbf{k}\right)$.
The scattering integral is given by (we drop the frequency argument
for the moment):\begin{widetext} 
\begin{eqnarray}
J_{\alpha\beta}\left(\mathbf{K}\right) & = & \int_{K^{\prime}Q}\delta\left(K-\left\vert \mathbf{K+Q}\right\vert +K^{\prime}-\left\vert \mathbf{K}^{\prime}\mathbf{-Q}\right\vert \right)F\left(\mathbf{K,K}^{\prime},\mathbf{Q}\right)\nonumber \\
 &  & \times\frac{e^{K}e^{K^{\prime}}}{\left(e^{\left\vert \mathbf{K}^{\prime}\mathbf{-Q}\right\vert }+1\right)\left(e^{K^{\prime}}+1\right)\left(e^{K}+1\right)\left(e^{\left\vert \mathbf{K+Q}\right\vert }+1\right)}\nonumber \\
 &  & \times\left(\psi_{\alpha\beta}\left(\mathbf{K+Q}\right)-\psi_{\alpha\beta}\left(\mathbf{K}^{\prime}\right)+\psi_{\alpha\beta}\left(\mathbf{K}^{\prime}\mathbf{-Q}\right)-\psi_{\alpha\beta}\left(\mathbf{K}\right)\right).
\end{eqnarray}
\end{widetext}Capital letters are used to denote dimensionless momenta,
i.e. $\mathbf{K}=\beta\hbar v\mathbf{k}$ etc. The Coulomb interaction
enters through the matrix element 
\begin{equation}
\alpha^{2}F\left(\mathbf{K,K}^{\prime},\mathbf{Q}\right)=F_{1}\left(\mathbf{K,Q\mathbf{-}K}^{\prime}\mathbf{,Q}\right)+F_{2}\left(\mathbf{K,K}^{\prime},\mathbf{Q}\right),
\end{equation}
where the functions $F_{i}$ are the $R_{i=1,2}$ defined in Ref.
\cite{Fritz2008}.

Next we formulate the solution of the Boltzmann equation as a variational
problem. The operator $\widehat{J}$ with 
\begin{equation}
\widehat{J}\left[\psi_{\alpha\beta}\right]\left(\mathbf{K}\right)_{\alpha\beta}=J_{\alpha\beta}\left(\mathbf{K}\right)
\end{equation}
is indeed self adjoint with respect to the scalar product 
\begin{equation}
\left\langle \varphi|\psi\right\rangle =\sum_{\alpha\beta}\int_{K}\varphi_{\alpha\beta}\left(\mathbf{K}\right)\psi_{\beta\alpha}\left(\mathbf{K}\right).\label{eq:Mode_scalar_product}
\end{equation}
If one uses that $F\left(\mathbf{K,K}^{\prime},\mathbf{Q}\right)$
is invariant under the substitution $\mathbf{K\rightarrow K}^{\prime}-\mathbf{Q}$
and $\mathbf{K\rightarrow K}^{\prime}+\mathbf{Q}$, one finds that
the solution of the Boltzmann equation can be obtained from the minimum,
i. e. $\frac{\delta Q\left[\psi\right]}{\delta\psi}=0$, of the functional
\begin{equation}
Q\left[\psi\right]=\frac{1}{2}\left\langle \psi\left\vert -2\pi\alpha^{2}\widehat{J}-i\Omega\frac{e^{K}}{\left(e^{K}+1\right)^{2}}\right\vert \psi\right\rangle +\left\langle S|\psi\right\rangle 
\end{equation}
with 
\begin{equation}
S_{\alpha\beta}\left(\mathbf{k}\right)=\frac{Ke^{K}}{\left(e^{K}+1\right)^{2}}I_{\alpha\beta}\left(\mathbf{k}\right).
\end{equation}

We now turn to the collision integral. It can be devided into two
parts: the so called collinear scattering part, where the momenta
of scattered particles are parallel, and the remaining scattering
processes. The former is dominant by a factor $\log{\alpha^{-1}}$,
where $\alpha$ is the coupling constant. Separating the operator
$\hat{J}$ into a part discribing only collinear scattering processes
(c) and the non-collinear part (nc) we write
\begin{equation}
\alpha^{2}\hat{J}=\alpha^{2}\log{\left(\alpha\right)^{-1}}\hat{j}_{c}+\alpha^{2}\hat{j}_{nc}.
\end{equation}
Assume that the collinear part projects $m$ so called zero modes
$\psi_{i}$, $i\in\left\{ 1,2,...m\right\} $onto zero, i.e.
\begin{equation}
\hat{j}_{c}\left[\psi_{\alpha\beta}^{i}\right]=0.
\end{equation}
We expand the function $\psi_{\lambda,\bm{k}}$ in eigenmodes of $\hat{j}_{c}$
with eigenvalues $b_{n}$:
\begin{equation}
\psi_{\alpha\beta}=\gamma_{0}\psi_{\alpha\beta}^{0}+...+\gamma_{m}\psi_{\alpha\beta}^{m}+\sum_{n>m}\gamma_{n}\psi_{\alpha\beta}^{n}.\label{eq:Inter_2}
\end{equation}
Let us abbreviate the left hand side of the Boltzmann equation (\ref{eq:Linearized_Boltzmann})
as $\mathscr{D}_{\alpha\beta}$
\begin{equation}
\mathscr{D}_{\alpha\beta}=\frac{e^{K}}{\left(e^{K}+1\right)^{2}}\left(-i\Omega\psi_{\alpha\beta}\left(K,\Omega\right)+KI_{\alpha\beta}\left(\mathbf{K}\right)\right)
\end{equation}
For the Boltzmann equation we then have
\begin{eqnarray}
\mathscr{D}_{\alpha\beta} & = & 2\pi\alpha^{2}\log{\alpha}^{^{-1}}\sum_{n>m}\gamma_{n}b_{n}\psi_{\alpha\beta}^{n}+2\pi\alpha^{2}\hat{j}_{nc}\left[\sum_{n}\gamma_{n}\psi_{\alpha\beta}^{n}\right]\nonumber \\
 & = & 2\pi\alpha^{2}\log{\alpha}^{^{-1}}\sum_{n>m}\gamma_{n}b_{n}\psi_{\alpha\beta}^{n}+\alpha^{2}C_{\alpha\beta}\,.\label{eq:Inter_1}
\end{eqnarray}
The operator $\hat{j}_{c}$ is hermitian with respect to the scalar
product (\ref{eq:Mode_scalar_product}). Thus, it's eigenfunctions
are orthogonal to each other. Taking the scalar product $\left\langle \mathscr{D}|\psi^{n}\right\rangle $
one has
\begin{equation}
\gamma_{n>m}=\frac{\left\langle \mathscr{D}|\psi^{n}\right\rangle -\alpha^{2}\left\langle C|\psi^{n}\right\rangle }{\alpha^{2}\log{\left(\alpha\right)^{-1}}b_{n}},
\end{equation}
so that in the expansion (\ref{eq:Inter_2}) all eigenfunctions with
$n>m$ are supressed by a factor $1/\log{\left(\alpha\right)^{-1}}$.
In a first approximation, we therefore retain the zero modes only.

It turns out \cite{Mueller2009} that the zero modes of the collinear
scattering operator are constant and linear in $\left|\mathbf{K}\right|$.
\begin{equation}
\psi_{\alpha\beta}^{0}=\psi_{0}\left(\Omega\right)I_{\alpha\beta}
\end{equation}
and 
\begin{equation}
\psi_{\alpha\beta}^{1}=\psi_{1}\left(\Omega\right)KI_{\alpha\beta}.
\end{equation}
The $\psi_{0}$ and $\psi_{1}$ at $\Omega=0$ correspond to the coefficients
$C_{0}$ and $C_{1}$ of Eq. (\ref{eq:Chapman_ansatz}) of the main
text. Thus, we obtain to leading order 
\begin{equation}
\psi\left(K,\Omega\right)=\left(\psi_{0}\left(\Omega\right)+\psi_{1}\left(\Omega\right)K\right)I_{\alpha\beta}\left(\mathbf{K}\right).
\end{equation}
We can determine the functional $Q$ within the space of the two basis
functions and obtain 
\begin{equation}
Q=\frac{1}{2}\sum_{a,b=0,1}\psi_{a}X_{ab}\psi_{b}+\sum_{a=0,1}\psi_{a}S_{a}
\end{equation}
with 
\begin{equation}
X_{ab}=\alpha_{0}^{2}R_{ab}-i\Omega r_{ab}
\end{equation}
and
\begin{equation}
R_{ab}=-2\pi J_{ab}
\end{equation}
and
\begin{equation}
r_{ab}=\left\langle \psi_{b}\left\vert i\Omega\frac{e^{K}}{\left(e^{K}+1\right)^{2}}\right\vert \psi_{b}\right\rangle ,
\end{equation}
where the indices $a,\,b$ label the matrix elements of the corresponding
operators in the $2\times2$ Hilbert space spanned by the modes $\psi_{0}$
and $\psi_{1}$. Once $X_{ab}$ and $S_{a}$ are known we obtain the
distribution function from the minimum of $Q$ as 
\begin{equation}
\psi_{a}=\sum_{b}\left(X^{-1}\right)_{ab}S_{b}.
\end{equation}
It holds 
\begin{eqnarray}
\sum_{a=0,1}\psi_{a}S_{a} & = & \left\langle S|\psi\right\rangle \nonumber \\
 & = & \sum_{\alpha\beta}\int_{K}S_{\alpha\beta}\left(\mathbf{K}\right)\psi_{\beta\alpha}\left(\mathbf{K}\right)\nonumber \\
 & = & \frac{\pi}{12}\psi_{0}+\frac{9\zeta\left(3\right)}{4\pi}\psi_{1},
\end{eqnarray}
which gives $S_{0}=\frac{\pi}{12}$ and $S_{1}=\frac{9\zeta\left(3\right)}{4\pi}$.
To determine $r_{ab}$ we start from\begin{widetext}
\begin{eqnarray}
\left\langle \psi\left\vert i\Omega\frac{e^{K}}{\left(e^{K}+1\right)^{2}}\right\vert \psi\right\rangle  & = & i\Omega\sum_{\alpha\beta}\int_{K}\psi_{\alpha\beta}\left(\mathbf{K}\right)\psi_{\beta\alpha}\left(\mathbf{K}\right)\frac{e^{K}}{\left(e^{K}+1\right)^{2}}\nonumber \\
 & = & i\Omega\left(\psi_{0}^{2}\frac{\log\left(2\right)}{2\pi}+2\psi_{0}\psi_{1}\frac{\pi}{12}+\psi_{1}^{2}\frac{9\zeta\left(3\right)}{4\pi}\right)
\end{eqnarray}
\end{widetext}which gives 
\begin{equation}
r_{ab}=\left(\begin{array}{cc}
\frac{\log\left(2\right)}{2\pi} & \frac{\pi}{12}\\
\frac{\pi}{12} & \frac{9\zeta\left(3\right)}{4\pi}
\end{array}\right)
\end{equation}
Finally for the analysis of the matrix $R_{ab}$ we have to analyze
the collision integral $\left\langle \psi\left\vert 2\pi\alpha_{0}^{2}\widehat{J}\right\vert \psi\right\rangle $.
This analysis can be done numerically and yields 
\begin{equation}
R_{ab}\simeq\left(\begin{array}{cc}
0.874 & 0.623\\
0.623 & 1.671
\end{array}\right)
\end{equation}
In order to determine the viscosity we then consider the relation
between the stress tensor 
\begin{equation}
\tau_{\alpha\beta}=N\hbar v\sum_{\lambda}\int\frac{d^{2}k}{\left(2\pi\right)^{2}}\frac{\lambda k_{\alpha}k_{\beta}}{k}f_{\mathbf{k}\lambda}\left(\mathbf{x},t\right)
\end{equation}
and the forcing $X_{\alpha\beta}$, which yields the shear viscosity
$\eta$ with $\tau_{\alpha\beta}=\eta X_{\alpha\beta}$. Inserting
Eq. (\ref{lin1}) for the distribution function yields at zero frequency:
\begin{eqnarray}
\eta & = & \frac{N\left(k_{B}T\right)^{2}}{8\pi\hbar v^{2}}\int_{0}^{\infty}dK\frac{K^{2}e^{K}}{\left(e^{K}+1\right)^{2}}\psi\left(K\right)\nonumber \\
 & = & \frac{N\left(k_{B}T\right)^{2}}{8\pi\hbar v^{2}}\left(\frac{\pi^{2}}{6}\psi_{0}\left(0\right)+\frac{9\zeta\left(3\right)}{2}\psi_{1}\left(0\right)\right)\nonumber \\
 & \simeq & 0.449\frac{N\left(k_{B}T\right)^{2}}{4\alpha_{0}^{2}\hbar v^{2}}.
\end{eqnarray}
This is the result given in Ref. \cite{Mueller2009}. For completeness,
we also give the expression at finite frequency: 
\begin{equation}
\eta\left(\omega\right)=\frac{N\left(k_{B}T\right)^{2}}{4\hbar v^{2}}\sum_{i=1}^{2}\frac{a_{i}k_{B}T}{-i\hbar\omega+\kappa_{i}\alpha^{2}k_{B}T}
\end{equation}
with $a_{1}=0.8598$, $\kappa_{1}=1.9150$ and $a_{2}=0.001159$,
$\kappa_{2}=21.182$. The fact that the viscosity is governed by a
sum of two Drude contributions is a consequence having two relevant
modes in our analysis. It is curious that the second mode is much
smaller in weight and contributes only $\sim10^{-4}$ to the static
viscosity. This second Drude contribution has a characteristic scattering
rate more than an order of magnitude larger than the first one. For
all practical purposes is the viscosity dominated by the first Drude
peak, which yields a characteristic scattering rate 
\begin{equation}
\hbar\tau_{{\rm e.e.}}^{-1}=\kappa_{1}\alpha^{2}k_{B}T.
\end{equation}
In our discussion we use this scattering rate for electron-electron
scattering. For comparison, the scattering rate that enters the conductivity
is given by $\hbar\tau_{{\rm e.e.}}^{-1}=3.646\alpha^{2}k_{B}T$ \cite{Fritz2008}.
We also note, that for finite frequencies the above derivations of
the slip lengths in the different limits are valid with the replacement
$\eta\rightarrow\eta\left(\omega\right)$.

\section{Flow around a circular obstacle: Solution on an infinite domain\label{sec:Flow_Infinite_Domain}}

The general solution in polar coordinates (r, $\theta$) to Eq. (\ref{eq:Oseen's_equation})
can be given in terms of modified Bessel functions of the first and
second kind, $I_{m}$ and $K_{m}$:\begin{widetext}
\begin{eqnarray}
q_{r} & = & -U\sum_{n=1}^{\infty}A_{n}\frac{\cos{\left(n\theta\right)}}{r^{n+1}}-\frac{1}{4}U\sum_{m=0}^{\infty}B_{m}\left(\frac{2}{kr}+\sum_{n=1}^{\infty}\Phi_{m,n}\left(kr\right)\cos\left(n\theta\right)\right)\nonumber \\
q_{\theta} & = & -U\sum_{n=1}^{\infty}A_{n}\frac{\sin{\left(n\theta\right)}}{r^{n+1}}-\frac{1}{4}U\sum_{m=0}^{\infty}\sum_{n=1}^{\infty}B_{m}\Psi_{m,n}\left(kr\right)\sin\left(n\theta\right),\label{eq:q_gen_sol}
\end{eqnarray}
\begin{eqnarray}
\Phi_{m,n}\left(kr\right) & = & \left(K_{m+1}+K_{m-1}\right)\left(I_{m-n}+I_{m+n}\right)+K_{m}\left(I_{m-n-1}+I_{m-n+1}+I_{m+n-1}+I_{m+n+1}\right),\nonumber \\
\Psi_{m,n}\left(kr\right) & = & \left(K_{m+1}-K_{m-1}\right)\left(I_{m-n}-I_{m+n}\right)+K_{m}\left(I_{m-n-1}-I_{m-n+1}-I_{m+n-1}+I_{m+n+1}\right),
\end{eqnarray}
\end{widetext} with $k=U/\left(2\nu\right)$, where $\nu$ is the
kinematic viscosity $\nu=\left(v^{2}/\tilde{w}_{0}\right)\eta$ \cite{Tomotika1950}.
The Bessel functions in the above equations have the argument $\left(kr\right)$.
The pressure is given by
\begin{equation}
p=U\frac{\tilde{w}_{0}}{v^{2}}\frac{\partial\phi}{\partial x},\label{eq:pressure_phi}
\end{equation}
with
\begin{equation}
\phi=UA_{0}\log{\left(r\right)}-U\sum_{n=1}^{\infty}\frac{A_{n}}{n}\frac{\cos{\left(n\theta\right)}}{r^{n}}.\label{eq:phi_solution}
\end{equation}
For details of the calculation, we refer to Ref. \cite{Tomotika1950}.
The Reynolds number of the problem is
\begin{equation}
R=\frac{Ud}{\nu}=4ka.
\end{equation}

In our case, the general boundary condition of Eq. (\ref{eq:General_boundary_condition})
reads 
\begin{eqnarray}
q_{r} & = & -U\cos\left(\theta\right)\nonumber \\
q_{\theta} & = & U\sin\left(\theta\right)+\zeta\frac{\partial q_{\theta}}{\partial r}.\label{eq:q_boundary_contitions}
\end{eqnarray}
Inserting the general solution (\ref{eq:q_gen_sol}) into the boundary
conditions (\ref{eq:q_boundary_contitions}) we derive an infinite
set of coupled equations for $A_{n}$, $B_{m}$. If the set is truncated
at some $m_{max}$ and $n_{max}=m_{max}+1$ the coefficients $A_{n\leq n_{max}}$,
$B_{m\leq m_{max}}$ are uniquely determined. The higher the Reynolds
number, the larger $m$, $n$ have to be considered. Here, we restrict
ourselves to $m=0$. While it might be important to include terms
with higher $m$, $n$ to describe the behavior near the obstacle,
the pressure far away is gouverned by the $m=0$ term, which decays
slowest (see Eqs. (\ref{eq:pressure_phi}) and (\ref{eq:phi_solution})).
For completeness, we give the coefficients $A_{0}$, $A_{1}$ and
$B_{0}$:\begin{widetext}
\begin{eqnarray}
A_{0} & = & -\frac{B_{0}}{2k}\label{eq:A0_B0}\\
A_{1} & = & a^{2}-\frac{1}{2}a^{2}B_{0}\left((I_{0}(ak)+I_{2}(ak)K_{0}(ak)+2I_{1}(ak)K_{1}(ak)\right)\label{eq:A1_B0}\\
B_{0} & = & \frac{2(a+\zeta)}{(a+\zeta)I_{0}(ak)K_{0}(ak)+(a+3\zeta)I_{1}(ak)K_{1}(ak)}.\label{eq:B0_eq}
\end{eqnarray}
\end{widetext}

\end{document}